\documentclass[a4paper, 11pt, oneside]{article} % A4 paper size, default 11pt font size and oneside for equal margins

\usepackage{cite}
\usepackage{amsmath,amssymb,amsfonts}
\usepackage{algorithmic}
\usepackage{graphicx}
\usepackage{textcomp}
\usepackage{xcolor}
\usepackage{fancyhdr}
\usepackage[hyphens]{url}

\usepackage[normalem]{ulem}
\usepackage{subfig}
\usepackage[keeplastbox]{flushend}
\usepackage{multirow}
\usepackage{colortbl}
\definecolor{mygray}{gray}{.88}

\newcommand{\tabincell}[2]{\begin{tabular}{@{}#1@{}}#2\end{tabular}}

\usepackage{epsfig}
\usepackage{authblk}
\begin{document} 

%%%%封面内容编辑%%%%
\begin{titlepage} % Suppresses headers and footers on the title page

	\centering % Centre everything on the title page
	
	\scshape % Use small caps for all text on the title page
	
	\vspace*{\baselineskip} % White space at the top of the page
	
	%------------------------------------------------
	%	Title
	%------------------------------------------------
	
	\rule{\textwidth}{1.6pt}\vspace*{-\baselineskip}\vspace*{2pt} % Thick horizontal rule
	\rule{\textwidth}{0.4pt} % Thin horizontal rule
	
	\vspace{0.75\baselineskip} % Whitespace above the title
	
	{\LARGE AIBench Scenario: Scenario-distilling AI Benchmarking\\} % Title
	
	\vspace{0.75\baselineskip} % Whitespace below the title
	
	\rule{\textwidth}{0.4pt}\vspace*{-\baselineskip}\vspace{3.2pt} % Thin horizontal rule
	\rule{\textwidth}{1.6pt} % Thick horizontal rule
	
	\vspace{2\baselineskip} % Whitespace after the title block
	
	%------------------------------------------------
	%	Subtitle
	%------------------------------------------------
	
	%Subtitle here % Subtitle or further description
	
	\vspace*{3\baselineskip} % Whitespace under the subtitle
	
	%------------------------------------------------
	%	Editor(s)
	%------------------------------------------------
	
% 	Edited By
    % Authors' contributions
	
 	\vspace{9\baselineskip} % Whitespace before the editors
	This paper has been accepted by The 30th International Conference on Parallel Architectures and Compilation Techniques (PACT 2021)
% 	Section~\ref{Introduction} was contributed by Jianfeng Zhan and Wanling Gao. Section~\ref{Methodology} was contributed by Jianfeng Zhan, Wanling Gao, Lei Wang and Fei Tang. Section~\ref{identify} was contributed by Wanling Gao, Fei Tang, Chuanxin Lan, Chunjie Luo, Jiahui Dai, Zheng Cao, Xingwang Xiong, Zihan Jiang, Tianshu Hao, Fanda Fan, Fan Zhang, Yunyou Huang, Jianan Chen, Mengjia Du, Rui Ren, Chen Zheng, Daoyi Zheng, Haoning Tang, Kunlin Zhan, Biao Wang, Defei Kong, Tong Wu, Minghe Yu, Chongkang Tan, Huan Li, Xinhui Tian, Yatao Li, Gang Lu, Junchao Shao, Zhenyu Wang, Xiaoyu Wang, and Hainan Ye. Section~\ref{evaluation} was contributed by Fei Tang, Wanling Gao, Chuanxin Lan, and Xu Wen. Section~\ref{section_conclusion} was contributed by Jianfeng Zhan.
	
% 	{\scshape\Large Fei Tang\\ Wanling Gao\\Jianfeng Zhan\\Chunxin Lan\\Xu Wen\\ Lei Wang\\ Chunjie Luo\\Jiahui Dai\\Zheng Cao\\et al.\\ }
	
	%{\scshape\Large Wanling Gao\\ Fei Tang\\ Lei Wang\\Jianfeng Zhan\\Chunxin Lan\\ Chunjie Luo\\Yunyou Huang\\Jiahui Dai\\Hainan Ye\\Zheng Cao\\Daoyi Zheng\\Haoning Tang\\Kent Zhan\\Biao Wang\\Defei Kong\\Shimin Gong\\Minghe Yu\\Chongkang Tan\\Yabin Huang\\Xinhui Tian\\Yatao Li\\Junchao Shao\\Xiaoyu Wang\\Zhenyu Wang\\ } % Editor list
	
% 	\vspace{0.5\baselineskip} % Whitespace below the editor list

	\vfill % Whitespace between editor names and publisher logo
	
	%------------------------------------------------
	%	Publisher
	%------------------------------------------------
	
	%\plogo % Publisher logo
	%\def\BUlogo{\epsfig{file=ICT.pdf,height=3cm}}
	%\includegraphics[scale=0.135]{ICT.pdf}
	\epsfig{file=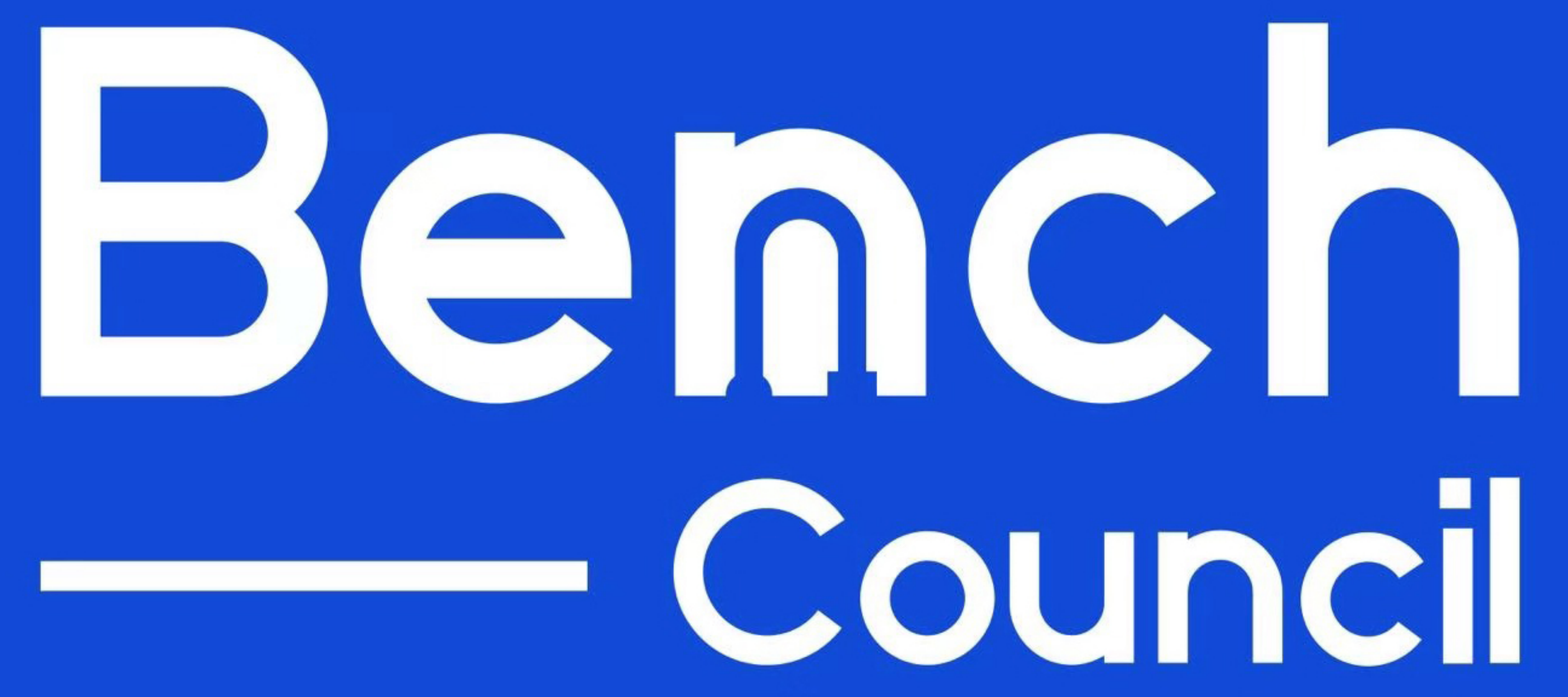,height=2cm}
	\textit{\\BenchCouncil: International Open Benchmark Council\\Chinese Academy of Sciences\\Beijing, China\\https://www.benchcouncil.org/aibench/scenario/} % Editor affiliation
	\vspace{1\baselineskip} % Whitespace under the publisher logo
 	
	Technical Report No. BenchCouncil-AIBench-Scenario-2021-1 % Publication year
	
	{\large August, 2021} % Publisher

\end{titlepage}

%----------------------------------------------------------------------------------------

%%%title here%%%
\title{AIBench Scenario: Scenario-distilling AI Benchmarking}

\author[1,2,3]{Wanling Gao}
\author[1,3]{Fei Tang}
\author[1,2,3]{Jianfeng Zhan\thanks{Jianfeng Zhan is the corresponding author.}}
\author[1,3]{Xu Wen}
\author[1,2,3]{Lei Wang}
\author[4]{Zheng Cao}
\author[1]{Chuanxin Lan}
\author[1,2,3]{Chunjie Luo}
\author[4]{Xiaoli Liu}
\author[1,3]{Zihan Jiang}

\affil[1]{State Key Laboratory of Computer Architecture, Institute of Computing Technology, Chinese Academy of Sciences \protect\\ \{gaowanling, tangfei, zhanjianfeng, wenxu, wanglei\_2011, lanchuanxin, luochunjie, jiangzihan\}@ict.ac.cn}
\affil[2]{BenchCouncil (International Open Benchmark Council)}
%\affil[3]{BenchCouncil R\&D Labs - Beijing, Guilin} 
\affil[3]{University of Chinese Academy of Sciences}
\affil[4]{Alibaba, \{zhengzhi.cz, sally.lxl\}@alibaba-inc.com}

\date{August, 2021}
\maketitle

%\newpage
\begin{abstract}

Modern real-world application scenarios like Internet services consist of a diversity of AI and non-AI modules with huge code sizes and long and complicated execution paths, which raises serious benchmarking or evaluating challenges. Using AI components or micro benchmarks alone can lead to error-prone conclusions.
This paper presents a methodology to attack the above challenge. We formalize a real-world application scenario as a Directed Acyclic Graph-based model and propose the rules to distill it into a permutation of essential AI and non-AI tasks, which we call a scenario benchmark. Together with seventeen industry partners, we extract nine typical scenario benchmarks. We design and implement an extensible, configurable, and flexible benchmark framework. We implement two Internet service AI scenario benchmarks based on the framework as proxies to two real-world application scenarios. 

We consider scenario, component, and micro benchmarks as three indispensable parts for evaluating. Our evaluation shows the advantage of our methodology against using component or micro AI benchmarks alone.
The specifications, source code~\footnote{Zenodo: https://doi.org/10.5281/zenodo.5158715\\GitHub: \url{https://github.com/BenchCouncil/aibench_scenario}}, testbed, and results are publicly available from   
\url{https://www.benchcouncil.org/aibench/scenario/}.

\end{abstract}

\section{Introduction}

Modern real-world application scenarios like Internet services are complex, which raises serious benchmarking and evaluating challenges.  First, they often adopt a microservice-based architecture consisting of a diversity of AI and non-AI modules with long and complicated execution paths across different datacenters.  Using  AI micro or component benchmarks alone can lead to error-prone conclusions. 

For Internet services, the overall system tail latency deteriorates even hundreds of times than that of a single AI component. For example,  for text classification--an AI component, the gap reaches up to 180x (Section~\ref{eva_server}).  
If we evaluate a new AI model or accelerator with text classification, it reduces the component tail latency. However, it contributes little to the overall system tail latency.
Note that tail latency represents the tail performance of a small fraction of requests, e.g.,  99th percentile latency, which is significant since these small fractions may exceed the median latency by orders of magnitude~\cite{li2014tales}.

Second, they dwarf the traditional workloads in terms of code sizes and deployment scales. For example, the traditional desktop workloads like data compression~\cite{speccpu2017}, image manipulation~\cite{speccpu2017}, are about one hundred thousand lines of code, running on a single node.  The Web server workloads~\cite{nginx} are hundreds of thousands of lines of code, running on a small-scale cluster, i.e., dozens of nodes. However, the modern AI  stacks (e.g., Spark~\cite{spark}, TensorFlow~\cite{abadi2016tensorflow}) alone are more than millions of lines of code, and the real-world workloads often run on a large-scale cluster, i.e., tens of thousands of nodes~\cite{barroso2009datacenter}. It is almost impossible to evaluate a real-world application scenario directly. Even it is possible, due to their huge code sizes, it will raise the fairness challenge in assuring the benchmark implementations' equivalence across different systems and the repeatability challenge in measurement errors.

 Third,  it is prohibitively costly for porting
a real-world application scenario
to an innovative system or architecture~\cite{gao2018bigdatabench, bailey1991parallel}. Especially, the real-world data sets, workloads, and even AI models are treated as first-class confidential
issues by their owners~\cite{hazelwood2018applied,ayers2018memory}.

While profiling has been applied to various aspects of benchmarking~\cite{conte1996accurate,alam2019zero}, we are not aware of evidence demonstrating it can holistically evaluate a large and complex system. 
For example, for Internet services, a few publicly available performance models or insights observed  through profiling~\cite{hazelwood2018applied,ayers2018memory,sriraman2020accelerometer,karandikar2020fireperf}
do not work across different systems or architectures. As there are no publicly available benchmarks, the state-of-the-art and state-of-the-practice are advanced only by the research staff within various service providers, which poses a considerable obstacle for our communities towards developing an open and mature research field.

\begin{figure}[tb]
\centering
\includegraphics[scale=0.35]{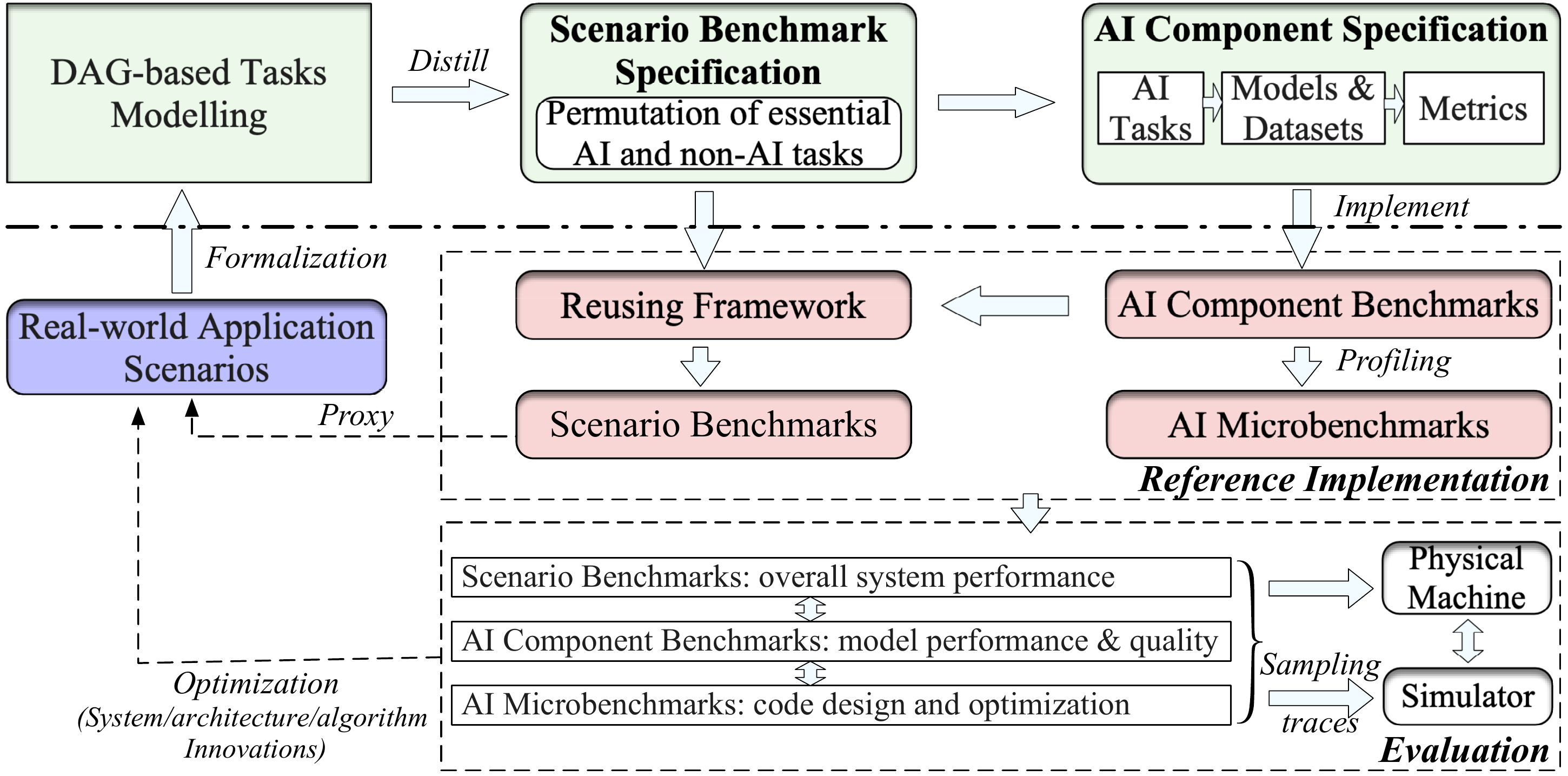}
\caption{Scenario-distilling Benchmarking.} 
\label{methodology1}
\end{figure}

This paper proposes a gray-box methodology that accesses no code but needs the industry's design input (Fig.~\ref{methodology1}), to attack the above challenge.
We formalize a real-world application scenario as a Directed Acyclic Graph-based model (in short, DAG model) and propose six rules to distill it into a permutation of essential AI and non-AI tasks as a  scenario benchmark. We call it a scenario-distilling methodology. 
Our methodology is to identify the critical path and primary modules of a real-world scenario since they consume the most system resources and are the core focuses for system design and optimization. 
We will extend to white-box (apply to industry users who can access the code) and black-box (without input) methodologies in the future.
In cooperation with seventeen industry partners, we extract nine vital scenario benchmarks. We design and implement a reusing benchmark framework, including the AI and
non-AI component libraries, data input, online inference, offline training, and deployment tool modules. Based on the framework, we implement two scenario benchmarks--E-commerce Search Intelligence and Online Translation Intelligence. With respect to its counterpart real-world application scenario, each scenario benchmark reduces the complexity by one or two orders of magnitude.  It is easier to achieve the latter's efficient implementation to avoid misleading evaluations using a suboptimal implementation. For a real-world application scenario, the complex software evolution also hinders high-performance implementation. 

We consider scenario, component, and micro benchmarks as three indispensable parts other than using component or micro benchmark alone.  A scenario benchmark lets the software and hardware designers obtain the overall system performance and determine the critical path's primary components. Furthermore, the system and architecture researchers use the AI component benchmarks that occupy the critical path to evaluate the AI model performance and quality targets. Finally, the code developers can use microbenchmarks obtained through profiling scenario benchmarks and component benchmarks to optimize the hotspot functions. They bridge a considerable gap from real-world application scenario deployment to simulator-based architecture research. 
The community can tune the system, architecture, and algorithm innovations on a  scenario benchmark without disclosing real-world application scenarios' confidentiality.   The developer can test the designs on a  scenario benchmark or obtain the trace of a scenario benchmark through sampling, a component benchmark, or a microbenchmark and feed them into simulators for earlier validation.  AIBench Scenario can use SimPoint~\cite{hamerly2005simpoint} or Pin~\cite{luk2005pin}.

Our evaluations demonstrate how to drill down from a scenario benchmark into the component benchmarks and zoom in on the microbenchmarks' hotspot functions.  
The evaluations on two CPU clusters and one GPU cluster show the advantage of our benchmarks against MLPerf and TailBench. 
We have several observations as follows:
(1) 
%The overall system performance of only one AI component varies significantly from that of putting this AI component in a scenario with a permutation of AI and non-AI components.
The performance characteristic of only one AI component varies from that of a scenario benchmark with a permutation of the previous AI component and other AI and non-AI components. The benchmark user must understand the implication of benchmarking using only one AI component like MLPerf in the context of a scenario benchmark. 
In serving the same requests for the online services, different AI components incur different latency. Some AI components contribute little to the overall system performance. (2) The overall system tail latency deteriorates
dozens of times or even hundreds of times over a single AI component, which we cannot predict using two state-of-the-art statistical models~\cite{delimitrou2018amdahl,urgaonkar2005analytical}. 
(3) We argue that on AI benchmarking, systems can benefit from analyzing both online services and offline AI training scenarios. Internet service architects generally attempt to balance the tradeoffs among service quality, input data dimensions, AI model complexity,  accuracy, and model update intervals.

We organize the rest of this paper as follows. Section 2 explains the motivation.
 Section 3 summarizes the related work. Section 4 presents how to distill and implement scenario benchmarks. 
Section 5 performs the evaluation.
Section 6 concludes.
\section{Motivation}\label{moti}

\subsection{Using component benchmarks alone may lead to error-prone conclusions. }\label{comperror}

Modern Internet services process millions of user queries daily; thus, the tail latency is important in terms of user experience~\cite{delimitrou2018amdahl}. However, a microservice-based architecture can contain various AI and non-AI modules, 
and consequently forms long and complicated execution paths. Previous AI benchmarking work provides a few micro or component benchmarks and fails to model an industry-scale application scenario's overall critical paths.

The overall system latency indicates that of the entire execution path, including AI and non-AI components.
Our experiments in Section~\ref{eva_server} show the overall system tail latency deteriorates dozens or even hundreds of times than a single component. The gap ranges from 2.2x – 180x between text classification and recommendation, respectively. When we evaluate a new AI model or accelerator using text classification, suppose it reduces the component tail latency. However, it still contributes little to the overall system tail latency.

Another case study is as follows.
Model pruning has been shown to be effective in improving inference speed~\cite{li2016pruning}. Evaluating with image-to-text a component of the Translation Intelligence scenario benchmark, a model-pruning implementation~\cite{li2016pruning} decreases the model size from 31MB to 21MB. It reduces the 99th percentile latency by  51\% from 4214 to 2081 milliseconds compared to the original model. The accuracy loss is 1\%, from 88\% decreasing to 87\%,  which may be considered acceptable considering the vast inference speedup. However, under a  scenario benchmark, the overall 99th percentile latency nearly remains the same (about 0.2\% reduction), with the same accuracy loss, which may be considered unacceptable as it reduces the service accuracy while providing no non-negligible computational benefit. Thus, scenario benchmarking is essential.

%\subsection{Can a Statistical Model Predict the Overall System Tail Latency?}

%{\color{blue}{Someone may argue a statistical model established through profiling the tail latency of many components can predict the overall system tail latency. Our answer is NO!}}

\subsection{Can a Statistical Model Predict the Overall System Tail Latency?}

The tail latency of many components has been studied by using various profiling-based statistical models. Recent work~\cite{marcus2021bao} proposes an algorithm to predict the tail latency of database query systems. However, for complex application scenarios like Internet services, the overall system tail latency prediction is not fully addressed.

In Section~\ref{statmodel}, we use two state-of-the-art queueing-based statistical models to estimate the overall system latency and tail latency: a simple queueing model~\cite{delimitrou2018amdahl} and a sophisticated queueing network model~\cite{urgaonkar2005analytical}. We find that the gaps between the actual average latency/tail latency and the theoretical ones are large through the experiments using two scenario benchmarks. For example, for E-commerce Intelligence, the average latency gap is 8.6x and 4.9x using the simple queueing model and the queueing network model, respectively. The 99th percentile latency gap is 3.3x using the simple queueing model. Due to the non-superposition property of the overall system tail latency--being not equal to the sum of each component--the 99th percentile latency 
is hard to be predicted accurately using a queueing network model~\cite{urgaonkar2005analytical}, so we do not report the 99th percentile latency gap using the queueing network model.

\subsection{Why Offline AI Training is also Essential in Scenario Benchmarking}

As our many industry partners have witnessed, they have to update the AI models in a real-time manner using an AI model for an online service. For example, one E-commence giant demands that the AI models are updated every one hour, and the updated model will bring in the award of about 3\% click-through rate and millions of profits. In Section~\ref{modelupdate}, our evaluation shows offline training should be included in scenario benchmarking, as it is essential to balance the tradeoffs among model update intervals, training overheads, and accuracy improvements.

\section{Related Work}

Several AI benchmark suites only provide component benchmarks, which may lead to error-prone conclusions.
Fathom~\cite{adolf2016fathom} provides eight deep learning component benchmarks implemented with TensorFlow, targeting six AI tasks.
DAWNBench~\cite{coleman2017dawnbench} is a component benchmark suite, which firstly pays attention to end-to-end performance--the training time to achieve state-of-the-art accuracy. 
MLPerf~\cite{mlperfweb,reddi2019mlperf,mattson2019mlperf} is an ML component benchmark suite targeting seven AI tasks and eight AI models for training, and six AI models for inference. Although MLPerf provides the inference for server mode, the server mode only contains one component and fails to cover a real-world industry-scale scenario's critical path. Our experiments illustrated in Section~\ref{comperror} also find that for complex scenarios, containing the execution path with only one component without the entire execution path is insufficient and may even lead to error-prone conclusions. 
MLPerf Tiny~\cite{banbury2021mlperf} is a deep learning benchmark for embedded and ultra-low-power devices.
TBD Suite~\cite{zhu2018tbd} is a component benchmark suite for DNN training, with eight neural network models for six AI tasks.  

Several AI benchmark suites only provide microbenchmarks, ignoring the quality target in benchmarking. 
DeepBench~\cite{deepbench} consists of four operations involved in training deep neural networks. %including three basic operations and recurrent layer operations. %Although the recurrent layer operation is the combination of several basic operations like convolution, while it is still simpler comparing to a component benchmarks.
DNNMark~\cite{dong2017dnnmark} is a GPU benchmark suite that consists of a collection of deep neural network primitives.% (micro benchmarks). % Both DeepBench and DNNMark 

%In total,  this benchmark suite only provides micro benchmarks, and lacks of the component and application benchmarks.

%BenchNN~\cite{chen2012benchnn} develops and evaluates software neural network implementations of 5 (out of 12) high-performance applications from the PARSEC Benchmark Suite. This benchmark only reimplement several PARSEC benchmarks using one software stack.

%It provides eight micro benchmarks while lacking of component and application benchmarks.
%Tonic Suite~\cite{hauswald2015djinn} presents seven neural network workloads that use the DjiNN service.

%Analogously, it only contains component-level benchmarks, and lacks of an end-to-end application benchmark that can depict the entire execution paths of industry scale application.

Additionally, 
Sirius~\cite{hauswald2015sirius} is an end-to-end IPA web-service application that receives voice and image queries and responds with natural language. TailBench~\cite{kasture2016tailbench} is a benchmark suite that consists of eight latency-critical workloads. CloudSuite~\cite{ferdman2011clearing} provides eight workloads for benchmarking cloud services.
MLModelScope~\cite{dakkak2020mlmodelscope} proposes a distributed platform for scalable model evaluation.
DeathStarBench~\cite{gan2019open} is a benchmark suite for microservice, including five workloads. They find a single poorly-configured microservice or slow server can degrade end-to-end latency by several orders of magnitude~\cite{gan2019open,gan2019seer}. 

AIBench distills and abstracts real-world application scenarios into AI Scenario, Training, Inference, Micro and Synthetic Benchmarks across Datacenter, HPC, IoT, and Edge~\cite{aibenchtutorial}. This paper focuses on scenario benchmark, others include AIBench Training~\cite{tang2021aibench}, HPC AI500~\cite{jiang2021hpc}, Edge AIBench~\cite{hao2018edge,hao2021edgeaibench}, and AIoTBench~\cite{luo2018iot}.

%As a marked departure from the previous benchmarking, we proposes an agile domain-specific benchmarking methodology speeding up the domain-specific software and hardware co-design. Following this methodology, twe have identify ten end-to-end application scenarios, distill sixteen representative AI tasks, Our evaluation shows AIBench It provides four integrated parts: end-to-end benchmarks discovering the overall system's effects;  the AI component benchmarks measuring the achieved performance and quality targets for sixteen representative AI tasks; fourteen micro benchmarks for fine tuning hot spot functions; and the reusing framework facilitates agile end-to-end benchmark building. 

\section{The Design and Implementation}~\label{des-imp}

Collaborating with the seventeen industry partners whose domains include search engine, e-commerce, social network, a news feed, video, etc., we distill their products or services into nine representative scenarios, as summarized in Table~\ref{application_scenario}. After identifying their primary components, we design and implement a reusing benchmark framework.
% based on their primary component libraries.
Finally, we implement two scenario benchmarks on the reusing framework: E-commerce Search Intelligence (in short, E-commerce Intelligence)  and Online Translation Intelligence (Translation Intelligence).

Subsection~\ref{distill_method} demonstrates how to distill a scenario into a high-level scenario benchmark specification. Subsection~\ref{implement_method} explains the general steps to implement a scenario benchmark on the reusing benchmark framework and the implementation details of these two scenario benchmarks.

\begin{table*}[htbp]
\scriptsize
\caption{Nine Representative Scenarios Extracted from the Seventeen Industry Partners.} \label{application_scenario}
\renewcommand\arraystretch{1.05}
%\footnotesize
\label{requirement}
\center %p{0.455in}|
\begin{tabular}{|p{0.6in}|p{1.1in}|p{0.8in}|p{0.6in}|p{0.5in}|p{0.5in}|}
\hline
 \textbf{Application Scenario} & \textbf{Involved AI Task} & \textbf{Involved Non-AI Task} & \textbf{Data} & \textbf{Metrics} & \textbf{Model Update  Frequency} \\

\hline
E-commerce Search Intelligence & Text/Image Classification; Learning to Rank;   Recommendation & Query Parsing, Database Operation, Indexing & User Data, Product Data, Query Data &  Precision, Recall, Latency & High\\
\hline
Online Translation Intelligence & Text-to-Text Translation; Speech Recognition; Image-to-Text & Query Parsing, Audio/Image Preprocessing & Text, Audio, Image & Accuracy, Latency & Low\\
%Language and Dialogue Translation & Text-to-Text Translation; Speech Recognition & Query Parsing; Audio Preprocessing & Text, Speech & Accuracy, Latency & Low\\
\hline
Content-based Image Retrieval & Object Detection; Classification; Spatial Transformer;  Image-to-Text &Query Parsing, Indexing & Image & Precision, Recall, Latency & High\\
\hline
Web Searching & Text Summarization;  Learning to Rank;   Recommendation &Query Parsing, Indexing, Crawler & Product Data, Query Data &  Precision, Recall, Latency & High\\
\hline
Facial Authentication and Payment & Face Embedding; 3D Face Recognition; & Encryption & Face Image & Accuracy, Latency & Low\\
\hline
News Feed & Recommendation & Database Operation, Basic Statistics, Filter & Text &   Precision, Recall & High\\
\hline 
%Photo Translation & Classification; Spatial Transformer; Image-to-Text Converter; Text-to-Text Translation& Query Parsing, Image Preprocessing & Image, Text &  Accuracy, BLEU, Latency & Low\\
%\hline
 Live Streaming  & Image Generation; Image-to-Image & Video Codec, Video Capture &Image & Latency & Low\\
\hline
 Video Services  & Image Compression; Video Prediction &  Video Codec &Video & Accuracy, Latency & Low \\
\hline
%Web map services & 3D object reconstruction & & & & \\
Online Gaming & 3D Object Reconstruction; Image Generation; Image-to-Image & Rendering & Image &  Latency &  Low\\ %Reinforce learning
\hline

\end{tabular}
\end{table*}

%A real-world application is complex, and we only distill the permutations of primary AI and non-AI tasks. Table~\ref{requirement} summarizes the list of important application scenarios. 

\subsection{The Distilling Methodology}\label{distill_method}

%As modern workloads like Internet services are not only diverse, but also fast changing and expanding, the traditional benchmark methodology that creates a new benchmark or proxy for every possible workload is prohibitively costly and even impossible~\cite{gao2018bigdatabench}. Hence, an agile scenario benchmarking methodology is extremely essential.  
%\subsubsection{Methodology.}

%Fig.~\ref{methodology1} summarizes our  methodology, including five steps as follows. % as the following five steps.

\begin{figure*}[tb]
\centering
\includegraphics[scale=0.32]{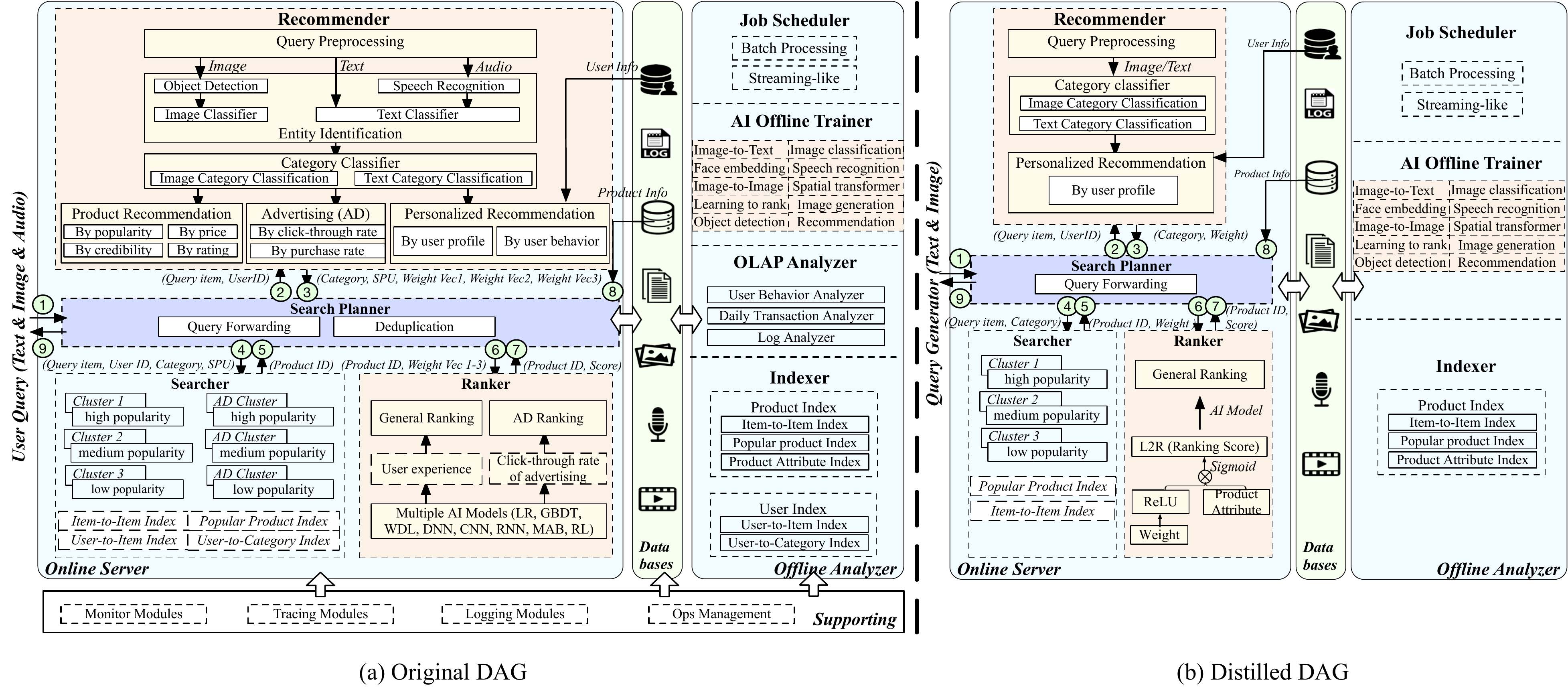}
\caption{The DAG Models of E-commerce Intelligence Before and After Distilling.} 
\label{ecommerce-arch}
\end{figure*}

We formalize a real-world application scenario as a DAG model, consisting of a series of tasks with dependencies. %Each task
Fig.~\ref{ecommerce-arch}(a) shows the DAG model of the real-world E-commerce Intelligence derived from one Internet service giant. % involves several modules and internal components. Choosing DAG is because of its expressiveness and wide use. 
%The input of this step is a candidate list of industry-scale real-world applications or future applications. For the former one, 
The formalization requires the domain knowledge, needing input from the primary industry partners. Our partners accept it as it does not expose confidential real-world workloads, data sets, and AI models. After the formalization, the high-level scenario specification is stated in a  paper-and-pencil approach and reasonably 
divorced from individual implementations~\cite{bailey1991parallel, zhan2019benchcouncil}.
%For future applications, the aim of the formalization is to define the full application requirement without considering the implementation details. 

%Second, the massive code size, extreme deployment scale, and complex execution path raise the repeatability and fairness challenges of benchmarking. For the future applications like autonomous driving, the formalization is a predefined DAG according to the application prospect.
%So the purpose of this step is to understand their essential components and the permutations of different components.

%Step One (Formalization). We investigate the important benchmarking requirements with the primary industry partners. The input of this step is a candidate list of industry-scale real-world applications. Just copying  the real-world applications is impossible for two reasons. First, they treat the real-world workloads, data sets, or models are confidential issues. Second, the massive code size, extreme deployment scale, and complex execution path raise the repeatability and fairness challenges of benchmarking. So the purpose of this step is to understand their essential components and the permutations of different components.

\begin{figure}[tb]
\centering
\includegraphics[scale=0.7]{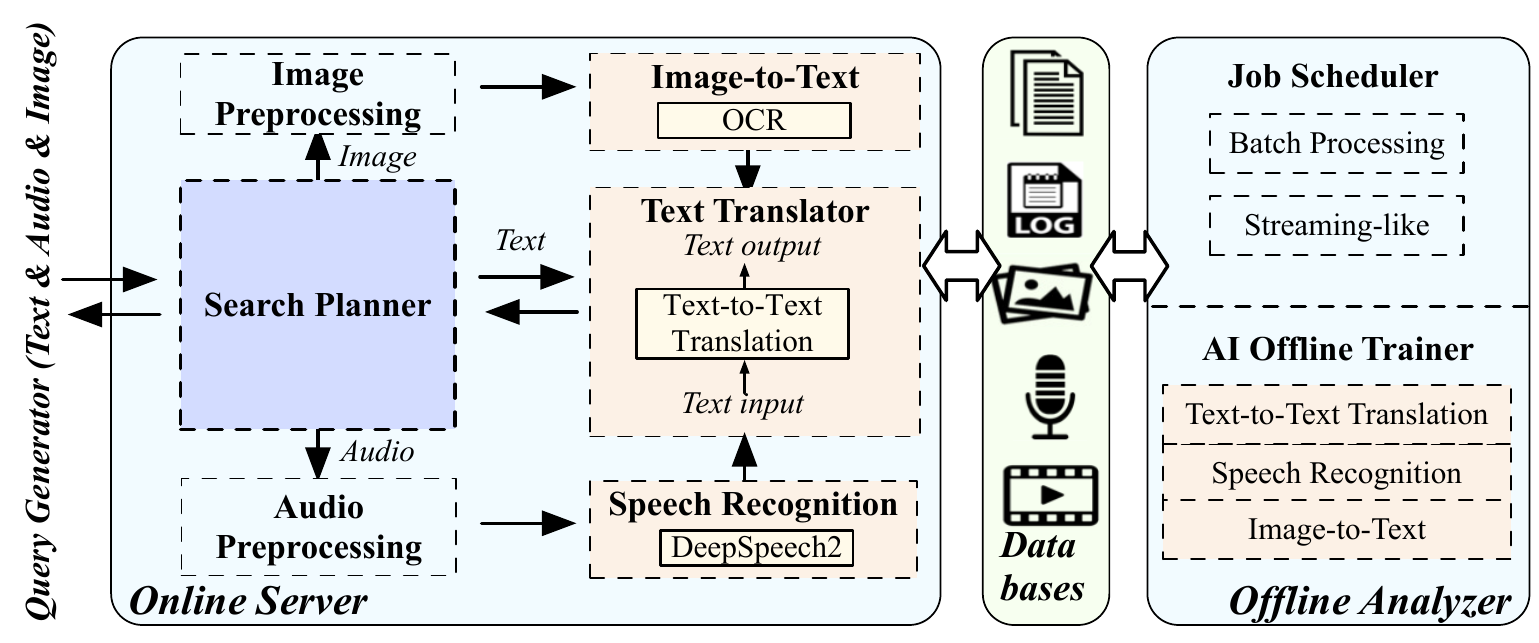}
\caption{The Distilled DAG Model of Translation Intelligence.} 
\label{translation-arch}
\end{figure}

We distill the scenario's DAG model into a permutation of essential AI and non-AI tasks and consider it a high-level scenario benchmark specification. %In nature, the scenario benchmark is a permutation of essential AI and non-AI tasks. 
 Fig.~\ref{ecommerce-arch}  shows the DAG models before and after distilling of E-commerce Intelligence. After distilling, we can reduce the complexity of the scenarios by one or two orders of magnitude.  
 Fig.~\ref{translation-arch} shows the DAG model of Translation Intelligence after distilling. Due to the page limitations, we only describe how we distill this scenario using text.
 To capture the essential parts of the original DAG, we propose six distilling rules.

%\begin{itemize}
%\item R1. The APP servers which represent different business units are not considered.
%\end{itemize}
%Comparing to the original one, we design a query generator to send queries to Online Server directly without APP server (R1) and only model the image and text searching, since the audio searching has a small percentage according to our industry partner (R2). 

\begin{itemize}
\item R1. We will keep only one essential branch among the DAG branches that have similar processing logic.
%will be duplicated to capture the essential parts. %Only one DAG branch is preserved if there are multiple branches having similar processing logic.
\end{itemize}

\emph{E-commerce Intelligence.} In the Recommender module, personalized recommendation, product recommendation, and advertising (AD) have similar processing logic though they consider different attributes and having other purposes. For example, personalized and product recommendations think user attributes and product attributes, respectively; recommendation cares about the user experience, while AD mainly concerns the click-through rate of advertised products. Here we only keep personalized recommendation as one essential branch.

\emph{Translation Intelligence.} In the Text Translator module, the translation microservices for different languages use a similar AI model while different training data. We only preserve the translation from English to Deutsch.
%We only consider the personalized recommendation, since the advertising and product recommendation have the similar processing logic with the personalized recommendation (R4). The difference is that they use different AI models and consider different purposes. For example, recommendation cares about the user experience, while advertising also concerns the click-through rate of advertising products.

\begin{itemize}
\item R2. We prune the DAG branch that occupies the smallest fraction that less than 1\%  in the whole scenario.
\end{itemize}

\emph{E-commerce Intelligence.} Within Recommender, there are three branches according to the query type---image, audio, and text. The audio query occupies a minuscule fraction and less than 1\%, so we prune it.
%only occupy XX\% according to our industry partner.

\emph{Translation Intelligence.} In the translation scenario, the fractions of text, image, and audio branches are more than 1\% according to our industry partner, so we reserve all three branches.

\begin{itemize}
\item R3. We remove the components that implement the auxiliary end-user functions.% irrelevant to the modelling scenario will be abandoned.
\end{itemize}

\emph{E-commerce Intelligence.} E-commerce Intelligence focuses on the product searching scenario, so we remove the OLAP analyzer that analyzes the product ordering and the supporting systems such as monitoring and logging. 

\emph{Translation Intelligence.} We also remove the supporting systems for tracing, monitoring, logging, etc.

\begin{itemize}
\item R4. For each AI component, we provide only one state-of-the-art or state-of-the-practice model for simplicity.
\end{itemize}

\emph{E-commerce Intelligence.} A scenario usually provides multiple AI models, which different businesses share. For simplicity, we only reserve one state-of-the-art or state-of-the-practice model for each AI component like Ranker.

\emph{Translation Intelligence.} We provide one state-of-the-art or state-of-the-practice model for its three AI components.

\begin{itemize}
\item R5. We combine the successive steps involved in similar components.
\end{itemize}

\emph{E-commerce Intelligence.} We combine the entity identification and category classifier within Recommender, as they both include image classifier and text classifier.

\emph{Translation Intelligence.} We combine the Text Translator, since all the three branches, i.e., the image, text, and audio, involve in text-to-text translation.

\begin{itemize}
\item R6. After performing R1 to R5, we will remove the components related to the pruned ones.
\end{itemize}

\emph{E-commerce Intelligence.} After pruning the AD branch in the recommender, we remove the AD cluster group in Searcher and AD ranking in Ranker accordingly. 
%Also, we remove the deduplication component in Search Planner, since the AD cluster group is removed, and there will be no repeated items returned by Searcher. Besides,  Indexer only builds the product indexes, since Searcher mainly uses the product attributes to search the product item.
Also, we remove the deduplication component in Search Planner since the AD cluster group is removed, and Searcher will return no repeated items. Besides,  Indexer only builds the product indexes since Searcher mainly uses the product attributes to search the product item.
%Also, we remove the deduplication component in Search Planner, as we only need the deduplication in searching the repeated items from the general product cluster and the AD cluster. Besides,  Indexer only builds the product indexes as Searcher focuses on the product attributes to search the product item.

\emph{Translation Intelligence.} After removing the translation microservices for the other languages, the related components are removed.

\subsection{How to Implement a Scenario Benchmark?}\label{implement_method}

\subsubsection{The Summary of Representative AI Tasks}\label{identify}

%We first introduction the problem domains identified from industries, including internet service, intelligence medicine, recognition science, etc. Then we illustrate a series of micro and application benchmarks that serve as a service or training model in AIBench framework. 

To cover a broad spectrum of AI Tasks, we thoroughly analyze the real-world application scenarios shown in Table~\ref{requirement}. We summarize the essential AI tasks and define the AI component specifications. Each AI component specifies an AI task description, a reference AI model, datasets, evaluation metrics, and state-of-the-art quality target~\cite{mattson2019mlperf}. In total, we identify sixteen representative AI tasks illustrated as follows. For each AI task, we implement the AI component benchmarks on both TensorFlow~\cite{abadi2016tensorflow} and PyTorch~\cite{pytorch}.% as . %Table~\ref{AIBench_component} summarizes the sixteen component benchmarks in AIBench.

%The sixteen AI tasks and component benchmarks are as follows. 
\emph{Classification} is to extract different thematic classes within the input data like an image~\cite{he2016deep} or text file~\cite{joulin2016fasttext}. %It is a typical task in Internet services or other application domains, and is widely used in multiple scenarios, like category prediction and spam detection.
\emph{Image Generation} is an unsupervised learning problem to mimic the distribution of data and generate images~\cite{arjovsky2017wasserstein}. %The typical scenario includes image resolution enhancement, which is used to generate a high-resolution image.
\emph{Text-to-Text Translation} needs to translate a text from one language to another~\cite{vaswani2017attention}. %, which is the most important field of computational linguistics. %It can be used to translate a search query and translate the dialogue.
\emph{Image-to-Text} is to extract the text information from an image, which is a process of optical character recognition (OCR)~\cite{wojna2017attention} or generate the description of an image automatically~\cite{vinyals2017show}. %It can be used to generate image caption or recognize the optical character.
\emph{Image-to-Image} is to convert an image from one representation to another one~\cite{zhu2017unpaired}, e.g., season change. %It can be used to synthesize the images with different facial ages and simulate virtual makeup. %Face aging can help search the facial images ranging different age stages.
\emph{Speech Recognition} is to recognize and translate a spoken language into text~\cite{amodei2016deep}. %This task is beneficial for voice search and voice dialogue translation.
\emph{Face Embedding} is to transform a facial image into a vector in an embedding space~\cite{schroff2015facenet}. %The typical scenarios are facial similarity analysis and face recognition.
\emph{3D Face Recognition} is to recognize the 3D facial information from multiple images from different angles~\cite{vieriu2015facial}. %This task mainly focuses on three-dimensional images and is beneficial to the facial similarity and facial authentication scenario.
\emph{Object Detection} is to detect the objects within an image~\cite{ren2015faster}. %The typical scenarios include vertical search and video object detection.
\emph{Recommendation} is to provide recommendations~\cite{he2017neural}. %This task is widely used for advertising recommendation, community recommendation, or product recommendation.
\emph{Video Prediction} is to predict the future video frames through predicting previous frames transformation~\cite{finn2016unsupervised}. %The typical scenarios are video compression and video encoding for efficient video storage and transmission.
\emph{Image Compression} is to compress the images and reduce the redundancy~\cite{toderici2017full}. %The task is vital for Internet services to minimize data storage overhead and improve data transmission efficiency.
\emph{3D Object Reconstruction} is to predict and reconstruct 3D objects~\cite{yan2016perspective}. %The typical scenarios are map search, light field rendering, virtual reality, and online gaming.
\emph{Text Summarization} is to generate a text  summary~\cite{nallapati2016abstractive}. %which is important for search results preview, headline generation, and keyword discovery.
\emph{Spatial Transformer} is to perform spatial transformations~\cite{jaderberg2015spatial} like stretching. %A typical scenario is space invariance image retrieval so that an image can be retrieved even if it is extremely stretched.
\emph{Learning to Rank} is to learn the attributes of a searched content and rank the scores for the results~\cite{tang2018ranking}. %which is the key for a search engine service.

Moreover, we profile all implemented component benchmarks and drill down into frequently appearing and time-consuming units of computation. Each of them constitutes a microbenchmark, which is easily portable to a new architecture and system and hence beneficial for fine-grained profiling and tuning.

\subsubsection{The Reusing Framework}

\begin{figure}[tb]
\centering
\includegraphics[scale=0.7]{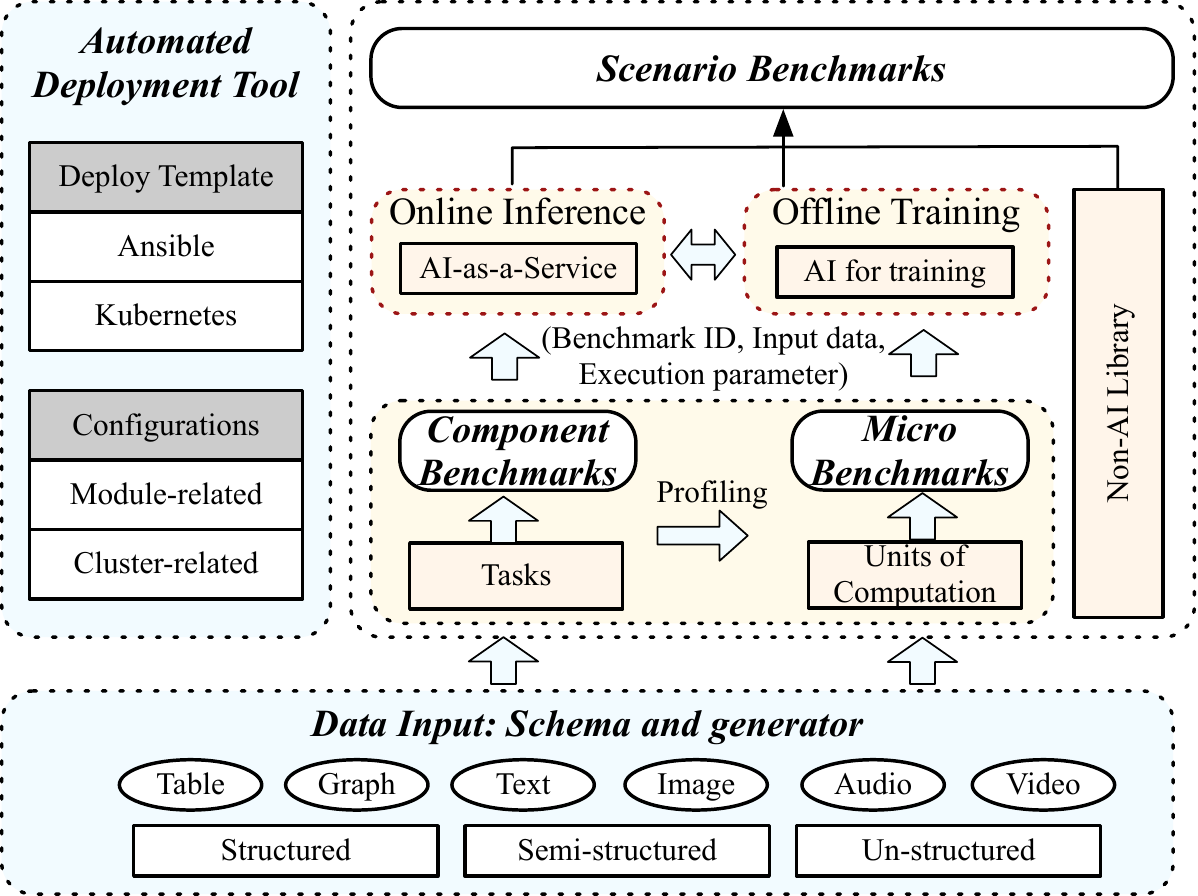}
\caption{The Reusing Framework.} 
\label{arch-design}
\end{figure}

As shown in Fig.~\ref{arch-design}, the reusing framework provides loosely coupled modules that are easily configured. It includes the data input,  offline training, online inference, non-AI libraries, and deployment tool modules. Based on the reusing framework, we can quickly implement a scenario benchmark. Since there are many other scenarios that we do not investigate, e.g., machine programming and event prediction, the framework may not suit all the scenarios. We will explore more scenarios and involved AI tasks to achieve broader coverage.
%as shown in Table~\ref{requirement}. 

% to constitute the entire lifecycle of an internet service application

The data input module is responsible for feeding data into the other modules. It collects representative real-world data sets from the authoritative public websites and our industry partners after anonymization.  We design the data schema to maintain the real-world data characteristics, alleviating the confidential issues.
 %Based on the data schema, a series of data generators are further provided to 
%generate large-scale user or product information. 
%The input module covers a wide spectrum of data characteristics, including structured, semi-structured, unstructured types, and diverse data sources like text, image, audio, and video. 
%To cover a wide spectrum of data characteristics, we take diverse data types (e.g., structured, semi-structured, unstructured), and different data sources (e.g., table, graph, text, image, audio, video) into account. 
Our framework covers a broad spectrum of data types and sources, integrates various open-source data storage systems, and supports large-scale data generation and deployment for partial data~\cite{ming2014bdgs}.

%The above identified sixteen AI tasks and corresponding component benchmarks serve as a component benchmark module in the framework.
%To achieve diversity and representativeness of our framework, we first identify prominent AI problem domains that play important roles in most important Internet services domains. And then we provide the concrete implementation of the AI algorithms targeting those AI problem domains as component benchmarks. 
% The component benchmarks and micro benchmarks are also a part of  the reusable  framework.  

We provide offline training and online inference modules to implement a scenario benchmark. First, the offline training module chooses one or more component benchmarks by specifying the input data and execution parameters like the batch size. Then the offline training module trains a model and provides the trained model to the online inference module. The online inference module loads the trained model onto the serving system, i.e., TensorFlow Serving~\cite{olston2017tensorflow}. The non-AI library module provides non-AI computation and database access, including query parsing, database operations, image and audio preprocessing, indexing, crawler, encryption, basic statistics, filter, video codec, video capture, rendering.
For a complex application scenario, the online inference, non-AI libraries, and offline training modules constitute an overall critical path together.%in the critical paths, an end-to-end benchmark is built.

The framework provides the deployment tools that contain two automated deployment templates using Ansible~\cite{hochstein2017ansible} and Kubernetes~\cite{hightower2017kubernetes} to support large-scale cluster deployments. The Ansible templates support scalable deployment on physical or virtual machines, while the Kubernetes templates are used to deploy on a container cluster. A configuration file needs to be specified for installation and deployment, including the module parameters like input data and the cluster parameters like nodes, memory, and network information. A user does not need to know how to install and run each module through the deployment tools.

\subsubsection{Design and Implementation of Scenario Benchmarks}~\label{imptrans}

Based on the reusing framework, we implement two scenario benchmarks---E-commerce Intelligence and Translation Intelligence. 
They model the complete use cases of a realistic E-commerce search and a real-world online translation scenario augmented with AI.%Among them, E-commerce Intelligence models the complete use case of a realistic E-commerce search augmented with AI, covering both text and image searching.
%Translation Intelligence benchmark models the complete use case of a realistic online translation scenario, covering text, image, and audio translation. 
%Both of them consist of four subsystems: Online Server, Offline Analyzer, Query Generator, and Data Storage, as shown in Fig.~\ref{ecommerce-arch}(b) and Fig.~\ref{translation-arch}, respectively.

The implementation steps are as follows. 
%\begin{itemize}
%\item 

(a) First, we choose the essential AI and non-AI component benchmarks and corresponding input data from the reusing framework according to Table~\ref{requirement}. 

%\item 
(b) Next, the permutation of component benchmarks is specified by the scenario benchmark specification. The permutations of E-commerce Intelligence and Translation Intelligence are shown in Fig.~\ref{ecommerce-arch}(b) and Fig.~\ref{translation-arch}, respectively. They consist of four subsystems: Online Server, Offline Analyzer, Query Generator, and Data Storage.
We will provide the permutation specifications of the other seven scenarios listed in Table~\ref{requirement}.

%\item 
(c) Then, we choose the deployment tools, i.e., Ansible and Kubernetes, and modify the reusing framework's deployment template. The deployment template includes the module-related configurations, i.e.,  input data, execution parameters, non-AI libraries, and cluster-related configurations, i.e., node, memory, and network information. Note that we have provided the default templates on our open-source website for both Ansible and Kubernetes. Users can change the configurations according to their environments.

%\item 
(d) Finally, we train the AI models of the selected AI component benchmarks using the offline training module and transfer the trained models to the online inference module.
Note that we also provide pre-trained models for direct use.

%\end{itemize}

\textbf{Implementation of E-commerce Intelligence.} Query generator simulates concurrent users and sends queries to Online Server based on a specific configuration. 
A query item provides either a text or an image to reflect the different search habits. The configuration designates the parameters like concurrency, query arriving rate, distribution, user thinking time, and ratios of text and image items. The configurations simulate different query characteristics and satisfy multiple generation strategies. Query Generator is based on JMeter~\cite{jmeter2017apache}.

Online Server receives query requests and performs searching and recommendation, integrating AI inference, consisting of four modules. 

%\begin{itemize}
%\item 
\emph{Search Planner} receives and forwards the queries, using the Spring Boot framework~\cite{webb2013spring}. 

%\item 
\emph{Recommender} is to analyze query items and provide a personalized recommendation, according to the user information obtained from User Database. It first conducts query preprocessing 
%spelling correction, and query rewriting 
and then predicts a query item belongs to which category based on two classification models---FastText~\cite{joulin2016fasttext} and  ResNet50~\cite{he2016deep}. FastText is used to classify a text query, while ResNet50~\cite{he2016deep} is used to classify an image query.
The Recommender module uses a deep neural network, proposed for e-commerce tasks~\cite{ni2018perceive}, to provide personalized recommendations. %It returns two vectors: the probability vector of the predicted categories and the user preference score vector of the product attributes, such as the user preferences for a brand, color, etc. 
We use TensorFlow Serving~\cite{olston2017tensorflow} to provide text classification, image classification, and online recommendation services.

%\item 
We deploy \emph{Searcher} on three different clusters on default, which follows an industry-scale deployment. %, to guarantee scalability and service efficiency. 
The clusters hold the inverted indexes of product information in the memory to guarantee high concurrency and low latency. According to the click-through rate and purchase rate, we classify the products into three categories according to popularity: high, medium, and low, and the proportion of data volume is 15\%, 50\%, and 50\%, respectively. Note that the high-popularity category is a subset of the medium-popularity one.  We store the indexes of products with different popularity in the other clusters.  

%Given a request, Searcher searches these three clusters one by one until the searched product data reach a threshold amount. Generally, the cluster that holds low-popularity products is rarely searched.
%So, for each category, Searcher adopts different deployment strategies. 
The cluster containing high-popularity product data has more nodes and more backups to guarantee searching efficiency. In comparison, the cluster containing low-popularity has the least number of nodes and backups. We use Elasticsearch~\cite{gormley2015elasticsearch} to set up and manage Searcher deploying on the three clusters.

%\item 
\emph{Ranker} uses the weight returned by \emph{Recommender} as initial weight and ranks the scores of products through a personalized L2R neural network~\cite{ni2018perceive}. Ranker uses TensorFlow Serving~\cite{olston2017tensorflow} to implement product ranking.

%\end{itemize}

Offline Analyzer performs job scheduling, AI model training, and updating. Job Scheduler includes batch and streaming-like processing, which updates the models every few hours or seconds accordingly. AI Offline Trainer trains learning models of chosen components or performs real-time model updates for online inference.

\textbf{Implementation of Translation Intelligence.} Query generator simulates concurrent users and sends either a text, an image, or audio to Online Server. % based on a specific configuration. 
The configuration information is the same as that of E-commerce Intelligence.

Online Server contains four modules. 
%\begin{itemize}
%\item 
\emph{Search Planner} is the entrance of the Online Server, which also uses the Spring Boot framework~\cite{webb2013spring}. 

%\item 
\emph{Image Converter} receives an image query and extracts the text information within the image. It first performs image preprocessing using a BASE64 image encoder since a RESTful API requires encoding binary inputs as Base64~\cite{restapi}. Then it converts the image into text using the Image-to-Text component---optical character recognition (OCR)~\cite{wojna2017attention}. 

%\item 
\emph{Audio Converter} receives an audio query and recognizes the text information within the audio. 
It first converts input audio into a WAV format with 16KHz and reuses Speech Recognition~\cite{amodei2016deep} component. 

%\item 
\emph{Text Translator} performs translations and reuses Text-to-Text Translation component~\cite{vaswani2017attention}. It receives text queries directly from Search Planner or the converted text data from Image Converter and Audio Converter.
%\end{itemize} 

Offline Analyzer performs job scheduling, which is similar to that of E-commerce Intelligence, except that they use different AI component benchmarks and update intervals. %For Translation Intelligence, AI Offline Trainer chooses three AI components from the reusing framework, including Speech Recognition, Image-to-Text (OCR), and Text-to-Text translation.

 %We are preparing the permutation templates of nine applications in Table~\ref{requirement}, and users can use them directly. For the other applications or future applications, users can design the permutation according to their requirements.

%The reusing framework provides two deployment tools, i.e., Ansible and Kubernetes. The deployment templates specify the deployment parameters and define the startup sequences of all the AI and non-AI components according to the permutation. The deployment parameters include the module-related configurations, i.e., input data path, execution parameters of component benchmarks, and cluster-related configurations, i.e., node, memory, and network information.  

%(3) Choose the deployment tools, i.e., Ansible and Kubernetes, and modify the deployment template. The template specifies the   permulation, execution parameters of component benchmarks, and cluster configuraton, e.g., node IP, memory, etc. We have provided the default templates on our open source web site. Users can change the configurations according to their own environments. Currently, we provide both Kubernetes and Ansible-based deployments.

%And then, we deploy the scenario benchmark and train the AI models of the selected AI component.

%The scenario benchmark is ready to be deployed using the selected deployment tool. It trains the AI models using the offline training module, and transfers the trained models to the online inference module. Note that we also provide pre-trained models for direct use.  %Finally provide services.

\subsubsection{Validation of a Scenario Benchmark}\label{validation}

Our methodology provides a two-level mechanism to verify whether a scenario benchmark can represent the real-world application scenario.

\textbf{Specification-level Validation.} On the specification level, it is easier to validate whether the scenario benchmark specification captures the essential parts of a scenario, as we can easily reach a consensus about the efficiency of the distilling rules with the industry partners' feedback. Our methodology and distilling rules only prune the auxiliary modules like the logging system and the non-critical path that occupies little weight; thus, from the specification level, the distilled DAG is close to the original DAG.%Otherwise, quantitative validation in terms of system and architecture characterization is almost impossible for two reasons. On one hand, the industry treats the implementation as confidential. On the other hand, due to their huge code sizes and complex software evolution, the implementations of the industry-scale scenarios are not necessarily the best ones.

\textbf{Implementation-level Validation.} On the implementation level, we compare the scenario benchmark with the real-world scenario from the perspectives of both overall system performance and single component performance. From the overall system performance perspective, we compare the overall system latency of the whole execution path under the same deployment. We define their deviation as the ratio of absolute latency difference between a scenario benchmark and a real-world one to the value of the real-world one—the lower the deviation, the closer the benchmark to reality. From the single component performance perspective, we replay the user queries and run the component on the same processor. Finally, we compare the latency of the component, system (e.g., CPU utilization), and micro-architectural (e.g., instructions per cycle) behaviors to check whether the scenario benchmark has similar behaviors with the real-world one.

\section{Evaluation}

This section summarizes our evaluation using the scenario benchmarks. 
Through the evaluations in Subsection~\ref{eva-end} and Subsection~\ref{modelupdate}, we demonstrate the advantages of scenario benchmarking in online services and offline training. We gain several insights that cannot be found using MLPerf~\cite{mlperfweb} and TailBench~\cite{kasture2016tailbench}.  
In Subsection~\ref{eva-end} and Subsection~\ref{drilldown}, we demonstrate how to drill down from a scenario benchmark into component benchmarks and zoom in on the hotspot functions of the microbenchmarks. The evaluations emphasize the necessity of scenario benchmarking and explain what benefits it can bring for the system and architecture communities.
%benchmarking scenario benchmarks to pinpoint the execution bottlenecks and provide optimization suggestions.

%In Subsection~\ref{eva-end}, we evaluate E-commerce search intelligence and explain why scenario-distilling AI benchmarking is necessary for both online server and offline trainer, and gain several insights, which can not be found using MLPerf~\cite{mlperfweb} and TailBench~\cite{kasture2016tailbench}. 
%In Subsection~\ref{eva-translate}, we evaluate another scenario benchmark---online translation intelligence. 
%In Subsection~\ref{{drilldown}}, we zoom in and drill down from scenario benchmarks to component benchmarks and until hotspot functions, emphasizing the necessity of benchmarking scenario benchmarks to pinpoint the execution bottlenecks and provide optimization suggestions.

%In Section~\ref{diverse_task}, we characterize diverse and distinct computation and memory patterns of sixteen AI tasks, emphasizing the necessity of including diverse AI tasks for benchmarking, which is also ignored by MLPerf~\cite{mlperfweb}. In Section~\ref{drillfunction},  we drill down to the hotspot functions, and analyze their execution stalls.
%, which is helpful for code optimization.

%and describe the effectiveness of AIBench for code developers and software or hardware designers.
%First, we evaluate the 

%evaluate the end-to-end AI application benchmarks introduced in Section~\ref{methodology}, including online server and all the sixteen AI component benchmarks.

\subsection{Experiment Setup}

%In this section, we describe the experimental setting and methodology. %environment for our evaluations using AIBench, and the method that we obtain the performance data.

%run a series of characterization experiments using AIBench to obtain insights for architectural studies and explore the impact of different technologies on datacenter computing, like virtualization technology. In this section, we present our experiment configurations and methodology on obtaining performance data.

\subsubsection{Node Configurations}

%Since the real-world industry-scale applications are usually deployed on clusters with multiple server types or even heterogeneous clusters, we prepare two CPU clusters with different CPU types and a CPU and GPU hybrid cluster.
%We deploy two CPU clusters and one CPU and GPU hybrid cluster for two scenario benchmarks.
Since CPU is widely used for inference in industry~\cite{hazelwood2018applied}, the online server of E-commerce Intelligence and Translation Intelligence is deployed on a 15-node CPU cluster and a 5-node CPU cluster, respectively. %, and a hybrid cluster consisting of 11-node CPU and 4-node GPU, for comparison. 
%The online server of Translation Intelligence is deployed on a 5-node CPU cluster. %, and a hybrid cluster consisting of 2-node CPU and 3-node GPU.
Their offline trainers are deployed on GPUs. Our another paper~\cite{tang2021aibench} has evaluated the AI offline training on other platforms like TPU.

For the 15-node CPU cluster, each node is equipped with two Xeon E5645 processors and 32 GB memory. For the 5-node CPU cluster, each node is equipped with two Xeon E5-2620 V3 (Haswell) processors and 64 GB memory. Each processor of the above two CPU clusters contains six physical out-of-order cores and disables Hyper-threading. They use the same OS version: Linux Ubuntu 18.04 with the Linux kernel version 4.15.0-91-generic, and the identical software versions: Python 3.6.8 and GCC 5.4. All the nodes are connected with a 1 Gb Ethernet network.

%The two CPU clusters include a 15-node CPU cluster and a 5-node CPU cluster. 
%For the 15-node CPU cluster, all the nodes are connected with a 1 Gb Ethernet network. Each CPU node is equipped with two Xeon E5645 processors and 32 GB memory. Each processor contains six physical out-of-order cores. Hyper-Threading is disabled. The OS version of each node is Linux Ubuntu 18.04 with the Linux kernel version 3.11.10. The software versions are Python 3.6.8 and GCC 5.4. %For comparison, we also deploy this scenario benchmark on a 11-node CPU and 4-node GPU cluster.

%To evaluate the performance on another platform, we deploy online translation intelligence on a 4-node CPU cluster. 
%For the 5-node CPU cluster, all the nodes are connected with a 1 Gb Ethernet network. Each CPU node is equipped with two Xeon E5-2620 V3 (Haswell) processors and 64 GB memory. Each processor contains six physical out-of-order cores. Hyper-Threading is disabled. The OS version of each node is Ubuntu 18.04. The Python and GCC versions are 3.6.8 and 5.4, respectively.

%For the two CPU and GPU hybrid clusters, the CPU node configurations are the same with that of corresponding scenario benchmark. 

Each GPU node is equipped with Nvidia Titan XP GPU. Every Titan XP owns 3840 Nvidia Cuda cores and 12 GB memory. The CUDA and Nvidia driver versions are 10.0 and 410.78, respectively. 
%The CPU and GPU hybrid cluster contains 11-node CPUs and 4-node GPUs. The CPU node configurations are the same with that of the 15-node CPU cluster. Each GPU node is equipped with Nvidia Titan XP GPU. Every Titan XP owns 3840 Nvidia Cuda cores and 12 GB memory. The CUDA and Nvidia driver versions are 10.0 and 410.78, respectively. 

\subsubsection{Benchmark Deployment}

%We deploy the online server of two AI scenario benchmarks introduced in Section~\ref{scenariotwo} on the above mentioned clusters. 

%The deployments of two AI scenario benchmarks introduced in Section~\ref{des-imp} are as follows.

\textbf{Online Server Settings of E-commerce Intelligence.} We deploy E-commerce Intelligence on the 15-node CPU cluster, containing one Query Generator node (Jmeter 5.1.1), one Search Planner node (SpringBoot 2.1.3), four Recommender nodes (TensorFlow Serving 1.14.0 for Text Classifier, Image Classifier, Recommendation, and Python 3.6.8 for Preprocessor), six searcher nodes (Elasticsearch 6.5.2~\cite{gormley2015elasticsearch}), one Ranker node (TensorFlow Serving 1.14.0), and two nodes for Data Storage (Neo4j 3.5.8 for User Database, Elasticsearch 6.5.2 for Product Database). %For comparison, we also deploy E-commerce Intelligence on the hybrid cluster consisting of 11-node CPU and 4-node GPU. The four AI related components are deployed on GPUs, including Text Classifier, Image Classifier, Recommendation, and Ranker. %Table~\ref{deployment} lists the benchmark deployments in detail.

\textbf{Online Server Settings of Translation Intelligence.} We deploy Translation Intelligence on the 5-node CPU cluster, containing one Query Generator node (Jmeter 5.1.1), one Search Planner node (SpringBoot 2.1.3), one Image Converter node (TensorFlow Serving 1.14.0 for Image-to-Text and Python 3.6.8 for Image Preprocessing), one Audio Converter node (TensorFlow Serving 1.14.0 for Speech Recognition and Python 3.6.8 for Audio  Preprocessing), and one Text Translator node (TensorFlow Serving 1.14.0). %Table~\ref{deployment} lists the benchmark deployment in detail.

\textbf{Offline AI Trainer Settings.} We deploy offline AI Trainers on 4-node GPUs for E-commerce Intelligence and on 3-node GPUs for Translation Intelligence to train AI models or update AI models in a real-time manner. %The GPU node contains the same GPU type like that of  the hybrid cluster---Nvidia Titan XP. 

\subsubsection{Performance Data Collection}

We use the network time protocol (NTP)~\cite{mills1985network} for synchronizing cluster-wide clock. % synchronization in all cluster nodes.
%and obtain the latency and tail latency metrics of the online server.
%A profiling tool---
We use Perf~\cite{de2010new} to collect the CPU micro-architectural data through the hardware performance monitoring counters (PMCs).
For GPU profiling, we use the Nvidia profiling toolkit---nvprof~\cite{nvprof} to track GPU performance. %To profile accuracy-ensured performance, we first adjust the parameters, e.g., batch size, to achieve the state-of-the-art quality target of that model on a given dataset, and then sample 1,000 epochs using the same parameter settings. For the GAN based model, whose accuracy is hard to measure, we set their parameters according to the referenced paper and reproduce the results.
We run each benchmark three times and report the average numbers.

\subsection{Benchmarking Online Services}~\label{eva-end}

%\subsection{The Necessity of End-to-end Benchmarking}\label{eva-end}

%{\color{blue}{
This subsection demonstrates how to drill down from a scenario benchmark into individual modules or even primary components for latency breakdown (Section~\ref{eva_server}), explains  why a statistical model cannot predict the overall system tail latency (Section~\ref{statmodel}), explores the factors impacting service quality (Section~\ref{eva-impact}), %illustrate why a scenario benchmark instead of component benchmarks is essential for system or architecture optimization (Section~\ref{errorprone}),}} 
and characterizes the micro-architectural behaviors (Section~\ref{microarch}).

\begin{figure}[tbp]
\renewcommand{\thesubfigure}{\arabic{subfigure}}
	\centering
	\subfloat[E-commerce Intelligence.]{\includegraphics[scale=0.35]{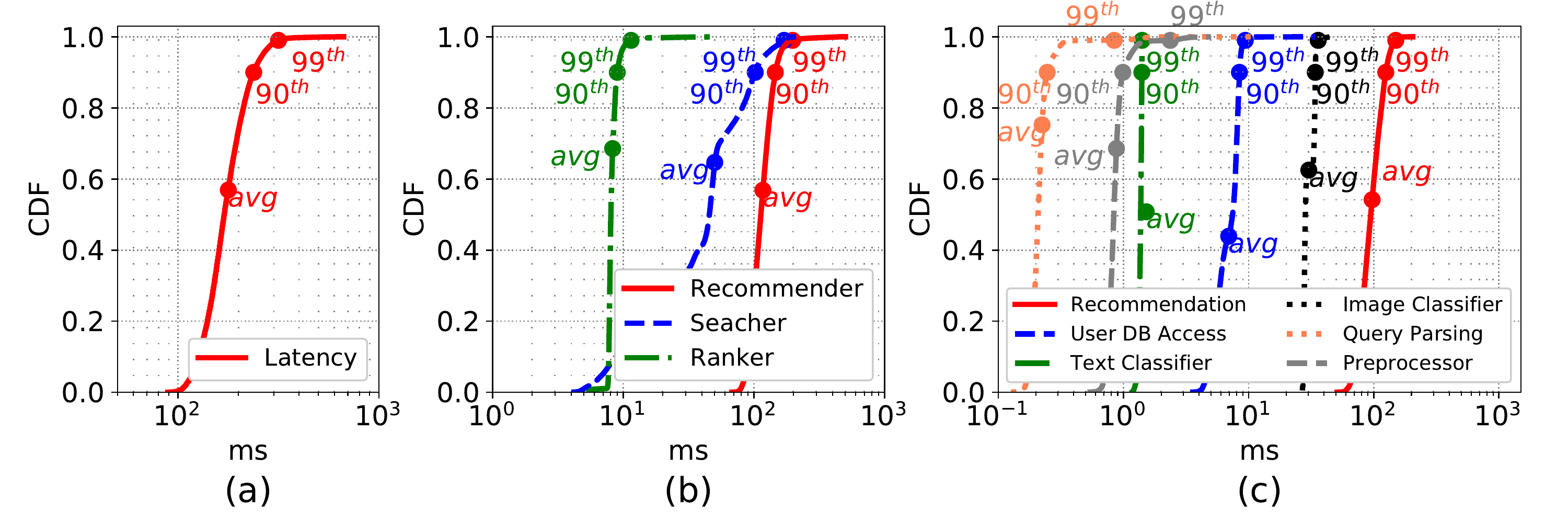}}\quad
	\subfloat[Translation Intelligence.]{\includegraphics[scale=0.35]{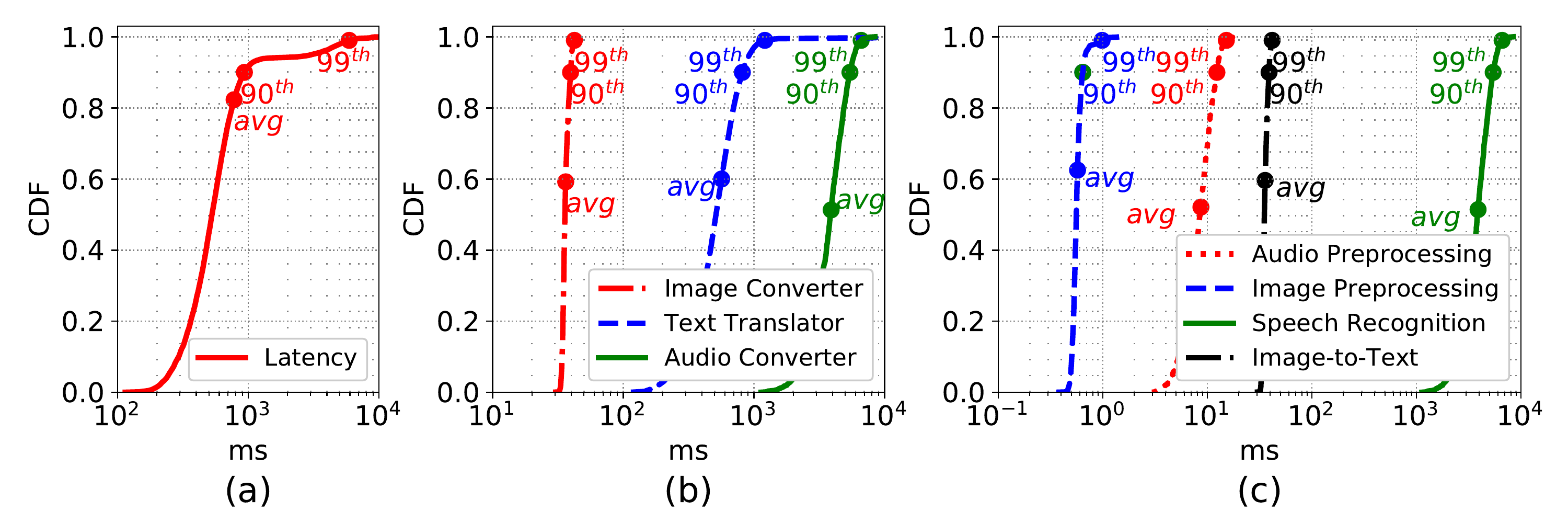}}\\	
	\caption{Overall System, Modules, and  Components Latency  Breakdown of  Two Scenario Benchmarks.}
	\label{latency:subfig} 
\end{figure}

%\begin{figure}
%\renewcommand{\thesubfigure}{\arabic{subfigure}}
%\centering
%\subfigure[E-commerce Intelligence.]{
%\label{latency:1} %% label for first subfigure
%\includegraphics[scale=0.28]{figures/latency-ecommerce1.eps}}
%\hspace{1in}
%\subfigure[Translation Intelligence.]{
%\label{latency:2} %% label for second subfigure
%\includegraphics[scale=0.28]{figures/latency-translation1.eps}}
%\caption{\color{blue}{Overall System, Modules, and  Components Latency  Breakdown of  Two Scenario Benchmarks.}}
%\label{latency:subfig} %% label for entire figure
%\end{figure}

%\begin{figure}[tb]
%\centering
%\includegraphics[scale=0.28]{figures/latency-ecommerce.eps}
%\caption{Latency of E-commerce Search Intelligence.}
%\label{latency}
%\end{figure}

%\textbf{The Latency and Critical Path of Online Server.}

%\subsubsection{Scenario Benchmarking is Necessary for Services}\label{eva_server}
\subsubsection{Latency Breakdown of Different Levels}\label{eva_server}

The latency is an essential metric to evaluate the service quality. This subsection demonstrates how to drill down into different levels for a detailed latency breakdown using two scenario benchmarks on the two CPU clusters.  The scenario benchmark configurations are as follows.

For E-commerce Intelligence, Product Database contains a hundred thousand products with 32-attribute fields. Query Generator simulates 2000 users with a 30-second warm-up time. A user sends query requests continuously every think time interval, following a Poisson distribution. According to history logs, we set the proportion of text and image queries as 99\% and 1\%, respectively. For Translation Intelligence, Query Generator simulates ten users with a 30-second warm-up time. The think time interval also follows a Poisson distribution. The proportion of text, image, and audio queries is 90\%, 5\%, and 5\%, respectively. We collect the performance numbers until 20,000 query requests have finished for both two scenario benchmarks. Also, we train each AI task to achieve the quality target of the referenced paper. 

Fig.~\ref{latency:subfig} shows the overall-system and individual-module latency of two scenario benchmarks, respectively.
%~\footnote{With respect to the real numbers in our industry partner, the number is quite high. They have taken many measures to decrease the overall latency.} 
%However, this baseline implementation indeed confirm the importance of the AI components in the critical paths as the data access and communication overhead can be further decreased.
%shows the whole latency of online server. 
From Fig.~\ref{latency:subfig}-1(a) and Fig.~\ref{latency:subfig}-2(a), we find the average, 90th percentile, and 99th percentile latency of the overall system of E-commerce Intelligence is 178, 238, and 316 milliseconds, respectively. Simultaneously, for Translation Intelligence, the number is 778.7, 934.4, and 5919.7 milliseconds, respectively. The two scenario benchmarks reflect different latency characteristics because of various permutations of AI and non-AI tasks.

We further give a latency breakdown of each module to identify the critical paths. Fig.~\ref{latency:subfig}-1(b) shows the latency of the Recommender, Searcher, Search Planner, and Ranker modules within E-commerce Intelligence. The latency of Search Planner is negligible, so we do not report it. We find that the Recommender occupies the most significant latency proportion: 117.06, 145.73, and 197.63 milliseconds for the average, 90th percentile, 99th percentile latency, respectively. The average, 90th percentile, 99th percentile latency is 50.12, 102.3,  170.21 milliseconds for Searcher, 8.3, 8.9, 11.5 milliseconds for Ranker, respectively.
 %For Ranker, the average latency is 8.29 milliseconds and the tail latency are no more than 12 milliseconds.  
Fig.~\ref{latency:subfig}-2(b) shows the latency of Image Converter, Audio Converter, and Text Translator modules within Translation Intelligence. Among them, Audio Converter incurs the most considerable latency: 3897.4, 5431.4, and 6613.3 milliseconds for the average, 90th percentile, 99th percentile latency, respectively. Text Translator also has a large proportion of latency, 565.8, 815.2, and 1212.1 milliseconds for the average, 90th percentile, and 99th percentile latency. Audio Converter's high latency is due to its high model complexity, spending thousands of milliseconds processing each audio query. %We further perform model pruning for audio converter to reduce its model complexity, however, the model accuracy reduces XX\%. Hence, there is a tradeoff between model pruning strategies and model accuracy.  
Text Translator needs to process all the queries eventually and results in query queueing, thus becoming the bottleneck of the overall system. 
We find that different AI components incur different latency for both scenario benchmarks, which their model complexity and dominance in the execution paths determine.

%Thus, AI benchmarking should be end-to-end to model the critical path.

Furthermore, Fig.~\ref{latency:subfig}-1(c) and Fig.~\ref{latency:subfig}-2(c) shows the latency breakdown of the most time-consuming modules: Recommender for E-commerce Intelligence, Audio Converter, and Image Converter for Translation Intelligence. Recommender includes
Query Parsing, Preprocessor, User DB Access, Image Classifier, Text Classifier, and Recommendation, as shown in Fig.~\ref{latency:subfig}-1(c). We find that Recommendation, Image Classifier (two AI components), and User DB Access (non-AI component) are the top three key components that impact the recommender's latency. The average latency of Recommendation (component) takes up 80\% of the average latency of Recommender (module) and occupies 54\% of the total latency of Online Server (overall system). The 99th percentile latency of Recommendation is 149.9 milliseconds, while the number of Recommender and Online Server is 197.63 and 316 milliseconds, respectively. For Text Classifier, its average latency and tail latency is less than two milliseconds, one-hundredth of the subsystem's latency. The overall system tail latency deteriorates dozens of times or even hundreds of times concerning a single component because of the following reasons. 
First, a single component may not be on the critical path. Second, even an AI component like Recommendation is critical; there are cascading interaction effects with the other AI  and non-AI components. 

From Fig.~\ref{latency:subfig}-2(c), we find that Speech Recognition has a significant impact on the latency, more than thousands of milliseconds both for the average and tail latency, due to its high model complexity. 

%Also, We use perf to sample the cache behaviors of the AI and non-AI components. We find that comparing to the non-AI  components, the AI components suffer from more cache misses in the memory hierarchy, especially L2 cache misses per Kilo instructions. The recommendation component suffers from 61 L2 cache misses per Kilo instructions, while the average number of non-AI components is only 37.
%\textbf{Recommendation component is the most important factor that influences the performance of online server}. %In the online server, Recommendation needs to run a forward inference using the recommendation model, and thus incurs larger latency.

We also analyze the execution time ratio of the AI components vs. non-AI components of the overall system latency. If we exclude the communication latency, the average time spent on the AI and the non-AI components is 137.12 and 58.16 milliseconds for E-commerce Intelligence. While for Translation Intelligence, nearly 99\% execution time is spent on the AI components. This observation indicates that the AI components' contributions to the overall system performance vary from different scenarios.
%indicating the AI components are essential critical path of the latter scenario.

\textbf{Reproducibility.} Reproducibility is significant for a benchmark. We repeat two scenario benchmarks five times to verify their reproducibility. We use the coefficient of variation (CV) as the metric, which is the standard deviation ratio to the mean value. %For E-commerce Intelligence, 
%the CV for the average, 90th percentile, and 99th percentile latency is 0.005, 0.007, and 0.014, respectively.
%%the standard deviations for the average, 90th percentile, and 99th percentile latency are 0.84, 1.64, and 4.3, respectively. We further compute the CV for the average, 90th percentile, and 99th percentile latency, which is 0.005, 0.007, and 0.014.
%For Translation Intelligence, the values are 0.006, 0.018, and 0.024, respectively. 
The CV for the average, 90th percentile, and 99th percentile latency is 0.005, 0.007, and 0.014 for E-commerce Intelligence, and 0.006, 0.018, and 0.024 for Translation Intelligence, respectively.
The experiments show that our scenario benchmarks have extremely low variance.

%\subsubsection{Micro-architectural Characterization}

\subsubsection{Validation of Scenario Benchmarks}

To validate how closely our scenario benchmark to real-world application scenarios, we take E-commerce Intelligence as an example to verify the effectiveness.

\begin{figure}[tb]
\centering
\includegraphics[scale=0.45]{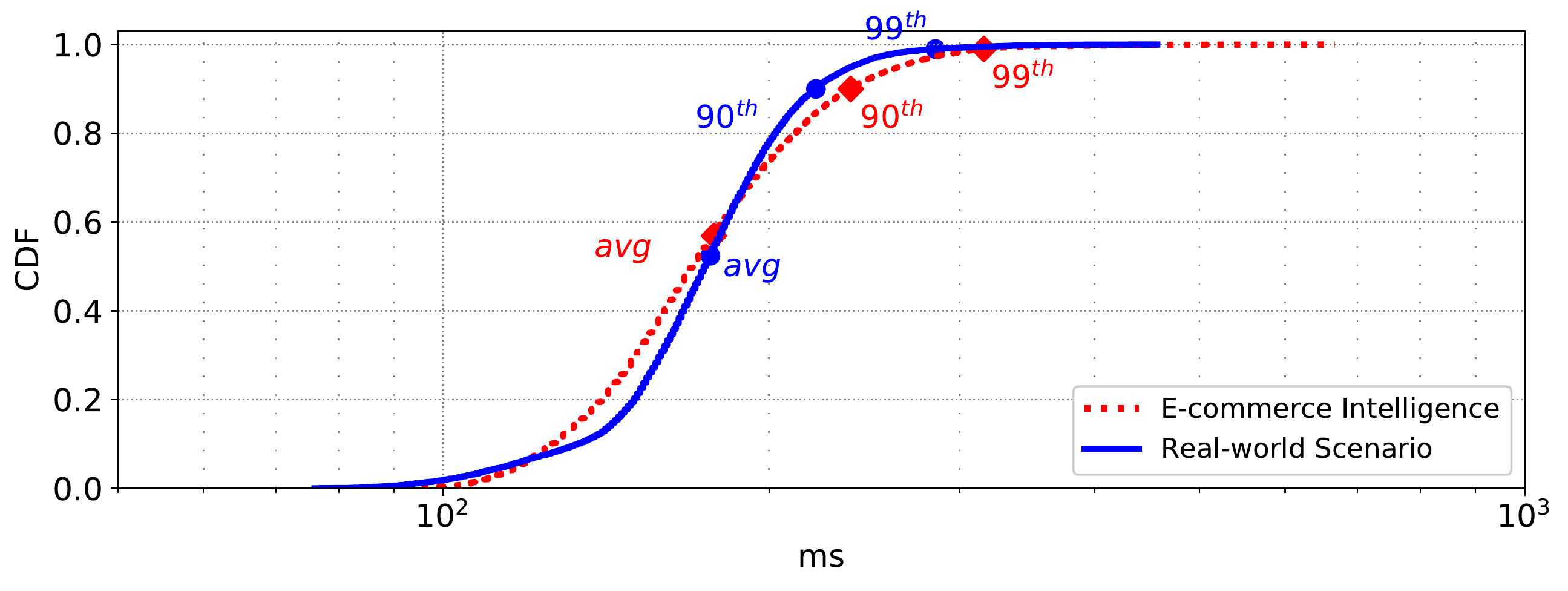}
\caption{Comparison of Overall System Latency between E-commerce Intelligence Benchmark and Real-world Scenario.} 
\label{real-comparison}
\end{figure}

We compare the E-commerce Intelligence scenario benchmark with the real-world E-commerce scenario from our industry partner. According to Section~\ref{validation}, we conduct the comparison from the overall system performance and single component performance. We deploy the real-world E-commerce scenario on our 15-node cluster and use the same configurations for the query generator. Fig.~\ref{real-comparison} shows the overall system latency of the whole execution path. We find that the average, 90th percentile and 99th percentile latency of the real-world scenario are 180.47, 225, and 299 milliseconds, respectively. The values are 178, 238, and 316 milliseconds respectively, for E-commerce Intelligence. The deviations for the average, 90th percentile, and 99th percentile latency are 1.4\%, 5.8\%, and 5.7\%, respectively, within the normal range. Hence, they share the consistent latency behaviors, demonstrating that the scenario benchmark captures the real-world E-commerce scenario's overall system performance.

We run the Recommendation component both within E-commerce Intelligence and real-world scenarios to compare a single component's performance. We choose Recommendation as it is the most time-consuming component and impacts the overall system latency.
 %Fig.~\ref{} shows the latency, system, and micro-architectural behaviors of two Recommendation components. 
The average latency of Recommendation for E-commerce Intelligence and real-world scenario are 97.03 and 103.68 milliseconds, respectively, and the deviation is 6\%. For the CPU utilization and instructions per cycle (IPC), the values are 51\% vs. 50\%, and 0.59 vs. 0.6, respectively. The deviations are 2\% and 1.7\% for CPU utilization and IPC. Thus, from a single component level, our benchmark still reflects its importance on the critical path and reflects similar workload behaviors within the real-world scenario.

%In conclusion, our methodology and scenario benchmarks can represent the real-world application scenarios, not only from the overall system perspective, but also from the single component perspective.

\subsubsection{Can a Statistical Model Predict the Overall System  Tail latency?}\label{statmodel}

As a scenario benchmark is more complicated than a microbenchmark or a component benchmark, we conduct experiments to check whether we can use a statistical model to predict the overall system tail latency.

The state-of-the-art work~\cite{delimitrou2018amdahl} uses the M/M/1 and M/M/K queueing models to estimate the p'th percentile latency. Since our experiments in Section~\ref{eva_server} only contain one Online Server instance, we adopt the M/M/1 model. %to predict the latency. %We deploy one instance of Online Server of both two scenario benchmarks. 
The p'th percentile latency ($T{p}$) and the average latency ($T{m}$) are calculated using the formulas:
$T{p}=-\frac{\ln \left ( 1-\frac{P}{100} \right )}{\mu - \lambda }$
, $T{m}=\frac{1}{\mu - \lambda }$.
$\mu$ is the service rate, following the exponential distribution. $\lambda$ is the arrival rate, following the Poisson distribution.

For E-commerce Intelligence, we obtain $\mu$ as 90 requests per second through experiments. Then we set  $\lambda$ as 3, 33, and 66 requests per second, respectively, to estimate 100, 1000, and 2000 concurrent users.
%Then we set  $\lambda$ as 3.0 requests per second (100 simulated users), 33 requests per second (1000 simulated users), and 66 requests per second (2000 simulated users), respectively. 
The average latency theoretical number is 11, 17, and 41 milliseconds for different settings, while the actual number is 141, 148, and 178 milliseconds, respectively. The average gap is 8.6x. The theoretical number of the 99th percentile latency is 52, 80, and 191 milliseconds, while the actual number is 261, 267, and 316 milliseconds, respectively. The average gap is 3.3x.
For Translation Intelligence, we get $\mu$ as 5.5 requests per second through experiments. 
Then we set  $\lambda$ as 0.3, 0.8, and 1.5 requests per second, respectively, to estimate 5, 10, and 20 concurrent users.
%Then we set  $\lambda$ as 0.30 requests per second (5 simulated users), 0.8 requests per second (10 simulated users), and 1.5 requests per second (20 simulated users), respectively. 
For different settings, the average latency's theoretical number is 192, 212, and 250 milliseconds, while the actual number is 788, 799, and 821 milliseconds, respectively. The average gap is 3.7x. The theoretical number of the 99th percentile latency is 885, 979, and 1151 milliseconds, while the actual number is 4283, 5354, and 5362 milliseconds, respectively. The average gap is 5.0x.

%We get the number of $\mu$---20 requests per second through the experiments. Then we set  $\lambda$ as 1.0 requests per second (10 simulated users), 9.1 requests per second (100 simulated users), and 16.7 requests per second (200 simulated users), respectively. For different settings, the theoretical number of the average latency is 53ms, 91ms, and 303ms, while the actual number is 123ms, 459ms, and 852ms, respectively. The average gap is 3.4 times. The theoretical number of the 99th percentile latency is 242ms, 422ms, and 1394ms, while  the actual number is 953ms, 5008ms, and 11980ms, respectively. The average gap is 8.1 times.  

The main reason for this vast gap is as follows. It is complex and uncertain about executing a scenario benchmark, and the service rate does not follow the exponential distribution. So, the M/M/1 model is far away from the realistic situation. 

%However, the more generalized model (e.g., G/G/1 model) is difficult to be used to calculate the tail latency. If we try to characterize the permutations of dozens of components in a scenario benchmark, we need a more sophisticated analytical model such as a queueing network model.

We further build a queueing network model~\cite{urgaonkar2005analytical} for  E-commerce Intelligence. The model consists of nine M/M/1 components, each of which corresponds with one component in E-commerce Intelligence, e.g., Text classifier in Fig.~\ref{ecommerce-arch}. We set $\lambda$ as 3, 33, and 66 requests per second, too. The average gap of the average latency between the theoretical and actual number is 4.9x. We do not report the tail latency gap because the queueing network is hard to accurately predict the overall tail latency due to its non-superposition property.

\subsubsection{Factors Impacting Service Quality}~\label{eva-impact}

We explore the impacts of the data dimension and model complexity on the service quality.

\begin{figure}[tb]
\centering
\includegraphics[scale=0.8]{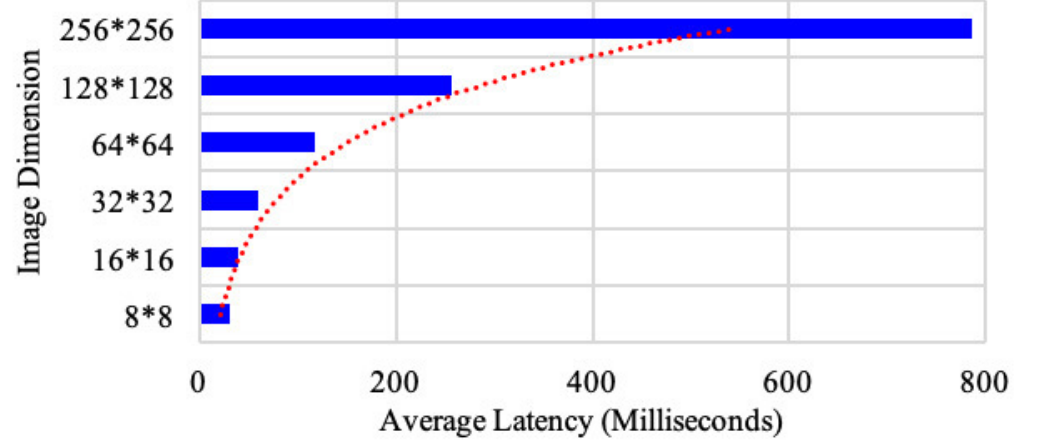}
\caption{Impact of Data Dimension on Service Quality.} 
\label{input-impact}
\end{figure}

\textbf{Impact of Data Dimension.} The data input has a significant impact on workload behaviors~\cite{gao2018motif,xie2018cvr}. We quantify its effect on the service quality through resizing the dimensions of input image data for Image Classifier in E-commerce Intelligence, including 8*8, 16*16, 32*32, 64*64, 128*128, and 256*256 dimensions. Fig.~\ref{input-impact} shows the average latency curve of the Image Classifier. We find that with the data complexity increases, the average latency deteriorates, but its slope gradually decreases. For example, the data dimension changes from 128*128 to 256*256, while the average latency deteriorates three times. The reason is that with the enlargement of the data dimension, the increase of continuous data accesses brings in better data locality. From the micro-architectural perspective, we notice that with the rise of data dimension, the IPC increases sharply from 0.34 to 2.37. The cache misses of all levels and pipeline stalls decrease, going down about ten or even dozens of times. Hence, resizing the input data dimension (data quality) to an appropriately larger one is beneficial for improving resource utilization. At the same time, there is a tradeoff between good service quality and enlarged data dimensions.

\textbf{Impact of Model Complexity.}
%\textbf{Tradeoff among Input Data Dimension and Service Quality}
%\textbf{Tradeoff among Service Quality, Model Accuracy, and Model Complexity.}
The online inference module needs to load the trained model and conducts a forward computation to obtain the result. Usually, increasing the depth of a neural network model may improve the model accuracy, but the side effect is that a larger model size results in longer inference time. For comparison, we replace ResNet50 with ResNet152 in the Image Classifier. The model accuracy improvement is 1.5\%, while the overall system 99th percentile latency deteriorates by 10x. %With ResNet152, the pipeline stalls increase two times, e.g., micro operation (uops) decode stall, reservation station full stall. 

Hence, Internet service architects should attempt to balance tradeoffs among data quality, service quality, model complexity, and model accuracy.

\subsubsection{Micro-architectural Characterization}~\label{microarch}

\begin{figure}[tb]
\centering
\includegraphics[scale=0.8]{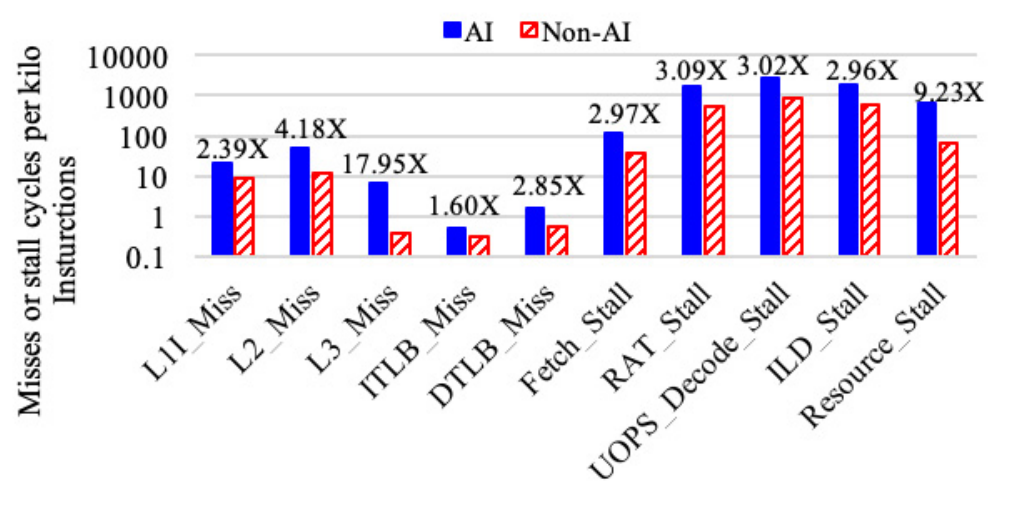}
\caption{The Cache and Pipeline Behavior Disparity between AI and non-AI Components. The y-axis indicates the average misses or stall cycles per kilo instructions of all AI and non-AI components; The labeled numbers reveal the AI components' ratios to Non-AI components.
} 
\label{cache-diff}
\end{figure}

We characterize the micro-architectural behaviors of both AI and non-AI components of two scenario benchmarks using Perf.
To compare AI and non-AI behaviors as a whole, we first report their average numbers of all AI and non-AI components.
%, with  being monitored. 
In our experiments, the AI components have lower IPC (instructions per cycle) (0.35) than that of the non-AI components (0.99) on average since they suffer from more cache misses, TLB misses, and execution stalls, even up to a dozen times. Fig.~\ref{cache-diff} shows the cache and pipeline behavior differences between AI and non-AI components.  
%We find that comparing to the non-AI components, AI components suffer from more cache misses and TLB misses, even up to a dozen times. 
The reason is that the AI components serving the services have more random memory accesses and worse data locality than non-AI components. Thus, they exhibit a larger working set and higher memory bandwidth.

For different AI components, we find in E-commerce Intelligence, the two components with higher latency, i.e., Image Classifier and Recommendation, suffer from higher backend bound while lower frontend bound than Text Classifier and Ranker.  The higher backend bound is mainly due to higher L1 data cache misses (more than 1.7x) for Image Classifier and more DRAM accesses (more than 1.4x) for Recommendation.  The higher frontend bound for Text Classifier and Ranker is mainly due to the higher frontend latency bound caused by large L1 instruction cache misses (more than 2.5x), and instruction TLB misses (more than 3.9x). %While for Ranker, the higher frontend bound is mainly due to higher frontend bandwidth bound caused by inefficient utilization of the Decoded Stream Buffer (DSB).
For Translation Intelligence, Image Converter and Audio Converter suffer from higher backend bound while lower frontend bound than those of Text Translator. The higher backend bound for Image Converter and Audio Converter is mainly due to higher L1 data cache misses (more than 2x) and more DRAM accesses (more than 1.5x). 
The lower frontend bound, incurred by inefficient utilization of the Decoded Stream Buffer (DSB), attributes to the lower frontend bandwidth bound (about two times lower).
%The higher frontend bound of Text Translator is mainly due to higher frontend bandwidth bound (more than 2 times), which is  mainly incurred by the inefficient utilization of the Decoded Stream Buffer (DSB).

\subsection{Benchmarking Offline Training}\label{modelupdate}

%Then we illustrate . % (Section~\ref{modelupdate}). 

Updating AI models in a real-time manner is a significant industrial concern in many scenarios shown in Table~\ref{requirement}. We evaluate how to balance the tradeoffs among model update intervals, training overheads, and accuracy improvements using AI Offline Trainer of E-commerce Intelligence on the Titan XP GPUs.% Taking E-commerce Intelligence as an example, we deploy AI Offline Trainer on four Titan XP GPUs.
%Offline trainer is deployed on the 4-node GPU of the hybrid cluster. 
%Each node trains a model. 
%Offline trainer is evaluated using the domain-specific metrics, i.e. real-time model update efficiency.

We resize the input data volume to investigate how to balance tradeoffs among model update intervals, training overheads, and accuracy improvements. 
For Image Classifier, using 60\% of training images, the training time is 2096.8 seconds to achieve the highest accuracy of 68.7\%. When using 80\% and 100\% of training images, the time spent is 2347.8 and 3170.7 seconds to achieve the highest accuracy of 71.8\% and 73.7\%, respectively. Hence, resizing the volume from 60\% to 100\%, 51\% additional training time brings in a 4.96\% accuracy improvement;  
From 80\% to 100\%, 35\% additional training time brings in a 1.9\% accuracy improvement. 
For Ranker, using 60\%, 80\%, and 100\% of training data, the training time is 839.3, 1222.8, and 1430,7 seconds, respectively. The highest accuracy is 12.4\%, 13.6\%, and 13.72\%, respectively, which is close to~\cite{tang2018ranking}. So, for this AI component, resizing from 60\% to 100\%, 70\% additional training time brings a 1.36\% accuracy improvement. Resizing from 80\% to 100\%, 17\% extra training time brings in a 0.12\% accuracy improvement.

%We conclude that offline training is essential for updating the online models in a real-time manner. 
We conclude that Internet service architects need to balance the tradeoffs among model update intervals, training overheads, and accuracy improvements. Moreover, for different AI components, the tradeoffs have subtle differences, and the evaluation should be performed using a scenario benchmarking methodology.  
%Hence, offline training should be integrated and evaluated is an integrated part of scenario benchmarking. It not only facilitates measuring the model update efficiency, but also provides a guidance on how to choose an optimal update interval to balance the tradeoff between training overhead and accuracy improvement.

%We adopt incremental learning method to update the models for online inference, and explore the relationship between the model update interval, training time overhead, and accuracy improvement. Our experiments show that comparing to the original training time and accuracy, 35\% additional training time brings in 1.9\% accuracy improvement for Image Classifier, and 10\% additional training time brings in 0.3\% accuracy improvement for Ranker. 

%Thus, offline training is an integrated part of scenario-distilling benchmarking. It not only facilitates measuring the model update efficiency, but also provides a guidance on how to choose an optimal update interval to balance the tradeoff between training overhead and accuracy improvement.

%\subsection{Evaluation on Online Translation Intelligence}
%
%\begin{figure}[tb]
%\centering
%\includegraphics[scale=0.28]{figures/latency-translation.eps}
%\caption{Latency of Online Translation Intelligence.} 
%\label{latency}
%\end{figure}

\subsection{Zooming in on Hotspot Functions}\label{drilldown}

%In this subsection, we demonstrate how to zoom in and drill down from scenario benchmark, to component benchmarks and until hotspot functions. 

The evaluations in Section~\ref{eva_server} demonstrate that drilling down from a scenario benchmark into the primary components lets us focus on the component benchmarks without losing the overall critical path. This subsection will further demonstrate how to zoom in on the hotspot functions of these primary component benchmarks.

Section~\ref{eva_server} shows the primary five AI components are on the critical path of the two scenario benchmarks and have significant impacts on the overall system latency and tail latency. They are Recommendation, Ranker, Image Classifier for E-commerce Intelligence, and Speech Recognition and Text Translator for Translation Intelligence. To zoom in on these component benchmarks' hotspot functions, 
we run both training and inference stages and use nvprof to trace the running time breakdown. To profile accuracy-ensured performance, we first adjust the parameters, e.g., batch size, achieve the state-of-the-art quality target of that model on a given dataset, and then sample 1,000 epochs of the training using the same parameter settings.

%The hotspot functions that occupy more than 1\% percentage of total running time are counted. %since there are dozens of function calls that take up less than 1\% percentage. 
Table~\ref{func-stat} shows the hotspot functions of the five primary components of the two scenario benchmarks. Note that we merge the time percentages of the same function and report the total sum. For example, the time percentage of Eigen::internal::EigenMetaKernel merges all the functions that call Eigen::internal::EigenMetaKernel with different parameters, like scalar\_sum\_op and scalar\_max\_op. %Each category contains a bunch of functions that solve the similar issue, and we report the total sum in Fig.~\ref{}. For example, convolution computation involves in multiple algorithms like winograd convolution.

\begin{table}[htbp]
\scriptsize
\caption{Hotspot Functions of Each AI Component.}
\renewcommand\arraystretch{1.2}
\scriptsize
\label{func-stat}
\center %p{0.455in}|
\begin{tabular}{|p{1.1in}|p{2.7in}|p{0.5in}|}
\hline
\textbf{Component} & \textbf{Function Name} & Time(\%) \ \\
\hline
\rowcolor{mygray} \multicolumn{3}{|l|}{AI Components of E-commerce Benchmark}\\ 
\hline
\multirow{4}{*}{\tabincell{l}{Recommendation}} & Eigen::internal::EigenMetaKernel & 39.6 \\
\cline{2-3}
& CUDA Memcpy & 23.4 \\
\cline{2-3}
& tensorflow::GatherOpKernel & 8.1 \\
\cline{2-3}
& tensorflow::scatter\_op\_gpu::ScatterOpCustomKernel & 4\\
%\cline{2-3}
%& cub::DeviceSelectSweepKernel & 4 \\
%\cline{2-3}
%& cub::DeviceReduceSingleTileKernel & 2.7 \\
\hline
\multirow{4}{*}{Ranker} &  Eigen::internal::EigenMetaKernel & 56.6\\
\cline{2-3}
& sgemm\_32x32x32\_NT\_vec & 9.1\\
\cline{2-3}
 & tensorflow::functor::ColumnReduceSimpleKernel & 8.7\\
 \cline{2-3}
 & sgemm\_32x32x32\_TN\_vec & 5 \\
%  \cline{2-3}
% & gemmk1\_kernel & 4.9 \\
 %\cline{2-3}
 %& tensorflow::BiasGradNHWC\_SharedAtomics & 2.9 \\
\hline
\multirow{5}{*}{\tabincell{l}{Image Classifier}} & Eigen::internal::EigenMetaKernel & 18.2 \\
\cline{2-3}
& maxwell\_scudnn\_128x128\_stridedB\_splitK\_interior\_nn & 15.6\\
\cline{2-3}
& maxwell\_scudnn\_128x128\_relu\_interior\_nn & 11.5 \\
\cline{2-3}
& maxwell\_scudnn\_128x128\_stridedB\_interior\_nn & 7.9\\
\cline{2-3}
& cudnn::detail::bn\_bw\_1C11\_kernel\_new & 7 \\
%\cline{2-3}
%& cudnn::detail::bn\_fw\_tr\_1C11\_kernel\_new & 4.3 \\
%\cline{2-3}
%& maxwell\_scudnn\_128x64\_stridedB\_splitK\_interior\_nn & 3.1 \\
%\cline{2-3}
%\hline
%\multirow{6}{*}{\tabincell{l}{Text \\Classifier}} & Eigen::internal::EigenMetaKernel & 75.9 \\
%\cline{2-3}
%& CUDA Memcpy & 19.6\\
%\cline{2-3}
%& tensorflow::UnsortedSegmentCustomKernel & 0.34 \\
%\cline{2-3}
%& sgemm\_32x32x32\_NT & 0.14\\
%\cline{2-3}
%& sgemm\_32x32x32\_NN & 0.12 \\
%\cline{2-3}
%& sgemm\_32x32x32\_TN & 0.12 \\
%\cline{2-3}
%& maxwell\_scudnn\_128x64\_stridedB\_splitK\_interior\_nn & 3.1 \\
%\cline{2-3}
\hline
\rowcolor{mygray} \multicolumn{3}{|l|}{AI Components of Translation Benchmark}\\ 
\hline
\multirow{4}{*}{\tabincell{l}{Text Translator}} &  Eigen::internal::EigenMetaKernel & 26.3 \\
\cline{2-3}
& sgemm\_128x128x8\_NN\_vec  & 17.6\\
\cline{2-3}
& sgemm\_128x128x8\_NT\_vec & 15.8\\
\cline{2-3}
& sgemm\_128x128x8\_TN\_vec & 8.3\\
%\cline{2-3}
%& maxwell\_sgemm\_128x64\_nt & 5.7 \\
%\cline{2-3}
%& maxwell\_sgemm\_128x128\_nt & 3.6 \\
%\cline{2-3}
%& maxwell\_sgemm\_128x64\_nn & 2.1 \\
%\cline{2-3}
%& tensorflow::BiasNHWCKernel & 1.9 \\
\hline
\multirow{4}{*}{\tabincell{l}{Speech Recognition}} & Eigen::internal::EigenMetaKernel   & 23.9\\
\cline{2-3}
& cudnn::detail::dgrad2d\_alg1\_1  & 17.4\\
\cline{2-3}
& cudnn::detail::dgrad\_engine & 13.4 \\
\cline{2-3}
& sgemm\_32x32x32\_NT\_vec & 9.4 \\
%\cline{2-3}
%& sgemm\_32x32x32\_NN\_vec & 7.7 \\
%\cline{2-3}
%& sgemm\_128x128x8\_TN\_vec & 6.5 \\
%\cline{2-3}
%& maxwell\_scudnn\_128x32\_stridedB\_splitK\_interior\_nn & 4.8 \\
%\cline{2-3}
%& CUDA Memcpy & 3.7 \\
%\cline{2-3}
%& maxwell\_scudnn\_128x32\_relu\_interior\_nn & 2.8 \\
\hline

%\multirow{6}{*}{\tabincell{l}{Image-to-\\Text}} & Eigen::internal::EigenMetaKernel   & 31.1\\
%\cline{2-3}
%& maxwell\_sgemmBatched\_64x64\_raggedMn\_nt  & 7\\
%\cline{2-3}
%& \tabincell{l}{tensorflow::functor::SwapDimension1And2InTensor-\\3UsingTile} & 6.8 \\
%\cline{2-3}
%& maxwell\_sgemmBatched\_64x64\_raggedMn\_nn & 4.1 \\
%\cline{2-3}
%& sgemm\_32x32x32\_NT\_vec & 3.8 \\
%\cline{2-3}
%& sgemm\_32x32x32\_NN\_vec & 3.5 \\
%\hline

\end{tabular}
\end{table}

\begin{figure}[tb]
\centering
\includegraphics[scale=0.7]{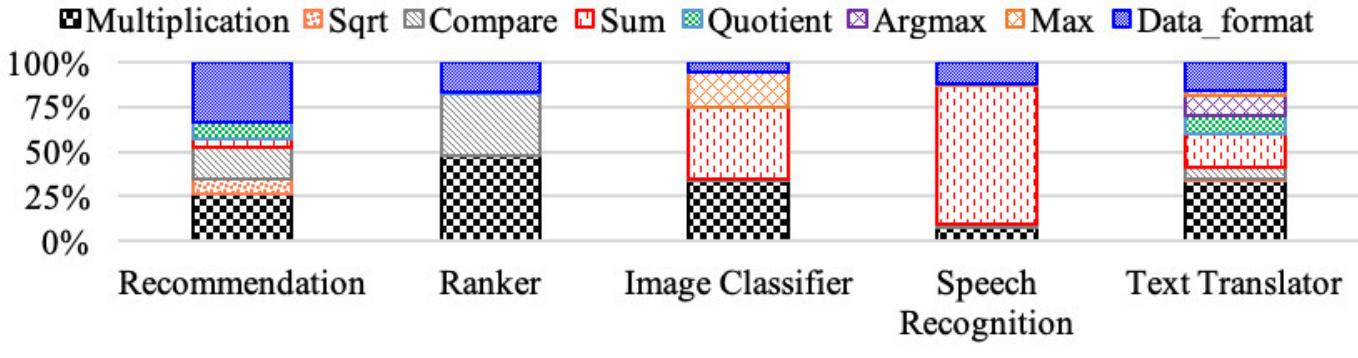}
\caption{Eigen::internal::EigenMetaKernel Runtime Breakdown—the most time-consuming function among the five AI components.} 
\label{eigen}
\end{figure}

%\textbf{Running time breakdown.} 
We find that Eigen::internal::EigenMetaKernel is the most time-consuming function among the five primary AI components. Eigen is a C++ template library for linear algebra~\cite{eigen} used by TensorFlow. Through statistics in Fig.~\ref{eigen}, we further find that within Eigen::internal::EigenMetaKernel, the most commonly used kernels are matrix multiplication, sqrt,  compare, sum, quotient, argmax, max, and data\_format computations. Here data\_format means variable assignment, data slice, data resize, etc.
%Further, for each micro benchmark, we summarize typical functions that occupy a large proportion of runtime among the important component benchmarks from scenario benchmarks, as shown in Table~\ref{}. 
We reveal that these five components and the corresponding hotspot functions are the optimization points for software stack and CUDA library optimizations and micro-architectural optimizations.

%\textbf{Stall Analysis.} 
Focusing on Eigen::internal::EigenMetaKernel, we further analyze their GPU stalls.
We find that the constant memory dependency stall, which immediate constant cache miss incurs, is the top one stall, occupying more than 50\% of total stalls on average. The memory dependency stall---a memory operation that cannot be performed due to the required resources not being available or fully utilized---occupies 20\% on average. 
The instruction fetch stall and execution dependency stall, which is due to an unready input required by the instruction, occupies about 14\% and 13\%, respectively. %Thus, for the Eigen functions, the top-1 stall is constant memory dependency stall.
%We analyze the percentages of eight kinds of stalls: instruction fetch stall which is the percentage of stalls because the next assembly instruction has not yet been fetched, execution dependency stall which is the percentage of stalls because an input required by the instruction is not yet available, memory dependency stall which is the percentage of stalls because a memory operation cannot be performed due to the required resources not being available or fully utilized, texture stall which is the percentage of stalls because of the under-utilization of the texture sub-system, synchronization stall which is the percentage of stalls due to a syncthreads call, constant memory dependency stall which is the percentage of stalls because of immediate constant cache miss, pipe busy stall is percentage of stalls because a compute operation cannot be performed because the compute pipeline is busy, memory throttle stall which is the percentage of stalls due to large pending memory operations.
%We find that for the above eight kinds of kernels, the mainly stalls are due to constant memory dependency stall, occupying more than 50\% of total stalls on average. Memory dependency stall, instruction fetch stall, and execution dependency stall occupies about 20\%, 14\%, and 13\% on average, respectively.
Possible optimization strategies include storing constants in global memory, improving data placement, and recomputing constant.

\section{Conclusion}

%In this paper, we argue component benchmarking is not sufficient, instead, end-to-end benchmarking is necessary for Internet service and other application scenarios AI.
This paper proposes a scenario-distilling methodology to tackle the challenges of benchmarking modern real-world application scenarios. We formalize a real-world application scenario as a DAG model and present the rules to distill it into the permutation of essential AI and non-AI tasks as a high-level scenario benchmark specification. Concerning the original scenarios, the scenario benchmarks reduce the complexity by one or two orders of magnitude. Together with seventeen industry partners, we extract nine scenarios and identify the primary components. We design and implement an extensible, configurable, and flexible benchmark framework. We implement two Internet service AI scenario benchmarks as proxies to two real-world application scenarios based on the framework.
 % among which we implement .  %
%, which is a proxy to a real-world application.
%This paper proposes a permutation of essential AI and non-AI tasks, which we call a scenario benchmark, as a proxy to a  scenario; 
We consider scenario, component, and micro benchmarks as three indispensable parts, and the evaluation shows the advantage of our methodology against using component or micro AI benchmarks alone. 

\section*{Acknowledgment}

We are grateful to the anonymous reviewers and our shepherd, Professor Justin Gottschlich. This work is supported by the Standardization Research Project of Chinese Academy of Sciences No.BZ201800001. Jianfeng Zhan is the corresponding author.

%\bibliographystyle{ieeetr}
%\bibliography{ref}

%\appendices
\section{Artifact Appendix}

%Submission and reviewing guidelines and methodology: \\
%{\em http://cTuning.org/ae/submission.html}

%%%%%%%%%%%%%%%%%%%%%%%%%%%%%%%%%%%%%%%%%%%%%%%%%%%%%%%%%%%%%%%%%%%%%
\subsection{Abstract}

{\em The artifact contains AIBench Scenario benchmarks, running and profiling scripts. It can support the characterization results in Chapter V.}

\subsection{Artifact check-list (meta-information)}

%{\em Obligatory. Use just a few informal keywords in all fields applicable to your artifacts and remove the rest. This information is needed to find appropriate reviewers and gradually unify artifact meta information in Digital Libraries.}

%{\small
\begin{itemize}
  \item {\bf Program:} AIBench Scenario benchmarks
  \item {\bf Compilation:} Python 3.6.8, Java 1.8.0
  \item {\bf Data set:} Anonymous data from industry or open-sourced data sets
  \item {\bf Run-time environment:} Ubuntu 18.04
  \item {\bf Hardware:} CPUs and GPUs
  \item {\bf Run-time state:} Disable Hyper-Threading
  \item {\bf Execution:} root user or users that can execute sudo without password
  \item {\bf Experiment:} Deploy the benchmarks and corresponding software stacks; run benchmarks; output the results
  \item {\bf Publicly available?:} Yes
  \item {\bf Workflow frameworks used?:} No
  \item {\bf Archived (provide DOI)?:} https://doi.org/10.5281/zenodo.5158715
\end{itemize}

%{\em Obligatory. Provide brief assessment of your artifacts
%based on {\em https://www.acm.org/publications/policies/artifact-review-badging}}:
%
%\begin{itemize}
%  \item {\bf Artifacts publicly available?: Yes}
%  \item {\bf Artifacts functional?:}
%  \item {\bf Artifacts reusable?:}
%  \item {\bf Results validated?:}
%\end{itemize}

%%%%%%%%%%%%%%%%%%%%%%%%%%%%%%%%%%%%%%%%%%%%%%%%%%%%%%%%%%%%%%%%%%%%%
\subsection{Description}

\subsubsection{How delivered}

%We have deployed an experiment environment with four software stacks, data motifs, and profiling scripts. You can access it using the commands:
%
%\textbf{\#ssh pact2018@159.226.41.249 -p 2233} with the passward \textbf{"pact"}.

The artifacts is available on Zenodo: https://doi.org/10.5281/zenodo.5158715, including all the source code scripts, and instructions. The Zenodo DOI is 10.5281/zenodo.5158715.
Also, the AIBench scenario benchmarks and framework are open sourced on GitHub: 

https://github.com/BenchCouncil/aibench\_scenario.

Note that the running or profiling scripts we provide suit for our cluster environment, like the node ip/hostname and port number, if you download and use it in your cluster environment, you need to modify the scripts.

\subsubsection{Hardware dependencies}

The AIBench scenario benchmarks can be run on all processors that can deploy Docker, Jmeter, SpringBoot, TensorFlow Serving, Elasticsearch and Neo4j stacks. However, for cache and pipeline behavior analysis, user need to find the performance counters corresponding to specific processor. We have provided profiling scripts for Xeon E5645 processor.

\subsubsection{Software dependencies}

Python 3.6.8, Java 1.8.0, Jmeter 5.1.1, SpringBoot 2.1.3, TensorFlow Serving 1.14.0, Elasticsearch 6.5.2, Neo4j 3.5.8, Docker 19.03 and Docker compose 1.29.

\subsubsection{Data sets}

We provide both anonymous data from industry and open-sourced data sets. Users can find where to download and how to use these data in the README file. 

%%%%%%%%%%%%%%%%%%%%%%%%%%%%%%%%%%%%%%%%%%%%%%%%%%%%%%%%%%%%%%%%%%%%%
\subsection{Installation}

{\em User need to install Python, Java, and Docker. For Jmeter, SpringBoot, TensorFlow Serving, Elasticsearch, and Neo4j, user can install them manually or use the deployment templates we provided. The install details can be found in the README file. }

%%%%%%%%%%%%%%%%%%%%%%%%%%%%%%%%%%%%%%%%%%%%%%%%%%%%%%%%%%%%%%%%%%%%%
\subsection{Experiment workflow}

Before profiling system and micro-architecture metrics of a scenario benchmark, users should make sure there is no other workload running.
Please refer to the README file within the artifacts package for detailed information.

\subsubsection{Deploy the benchmarks.}

We provide deployment scripts for two scenario benchmarks.

\textbf{Deploy E-commerce Intelligence:}

i. Under the following directory 

\#cd aibench\_scenario/benchmarks/ecommerce\_search

ii. Deploy benchmark: \#docker-compose up

\textbf{Deploy Translation Intelligence:}

i. Under the following directory

\#cd aibench\_scenario/benchmarks/translation

ii. Deploy benchmark: \#docker-compose up

\subsubsection{Run the benchmarks.}\label{run_benchmarks}

We provide running scripts for two scenario benchmarks. 

\textbf{Run E-commerce Intelligence:}

i. Load the data

\#cd aibench\_scenario/framework/online/non\_AI\_module/

benchmark-1.0-SNAPSHOT

\#./create\_index.sh 

\#./load\_data.sh

Note that you need to modify the IP address of the above two shell scripts according to your cluster configurations.

ii. Run the benchmark 

\#apache-jmeter-5.1.1/bin/jmeter -n -t aibench\_scenario/

framework/online/non\_AI\_module/load\_generator/

search\_loop\_2W.jmx -l result.txt -e -o report

Note that you need to modify the search\_loop\_2W.jmx file. 
First, use docker command to check the IP address of Search Planner;

\#docker ps

\#docker inspect \$ID 

The ID is the output of the command "docker ps" named Search Planner, the output will include the IP address of Search Planner.

Then change all the IP address of "172.18.11.91" in the search\_loop\_2W.jmx file to the actual IP address of Search Planner.

iii. Check the results

After the Step ii, it will generate a directory named "report". Within the directory, there is a file named "index.html", which can show the average latency, 90th percentile latency, and 99th percentile latency.
We also provide "run\_search.sh" to run the e-commerce intelligence benchmark. The user need to modify the scripts according to their own cluster environment.

\textbf{Run Translation Intelligence:}

i. Run the benchmark

\#cd aibench\_scenario 

\#apache-jmeter-5.2.1/bin/jmeter -n -t apache-jmeter-5.2.1/translation.jmx -l result.txt -e -o report 

Note that you need to modify the translation.jmx file.
First, use docker command to check the IP address of Translation Planner;
 
\#docker ps 

\#docker inspect \$ID 

The ID is the output of the command "docker ps" named Translation Planner, the output will include the IP address of Translation Planner.

Then change all the IP address of "172.18.13.26" in the translation.jmx file to the actual IP address of Translation Planner.
The result.txt will log the timestamp of all the queries.

ii. Check the results

After the Step i, it will generate a directory named "report". Within the directory, there is a file named "index.html", which can show the average latency, 90th percentile latency, and 99th percentile latency.

% iii. Choose one Spark motif: \#./run-spark.sh motif datasize

% Note that datasize parameter can be "large", "medium" or "small", means using large/medium/small data size,respectively.
% For example: \#./run-spark.sh graph large

% The sampling results of system and micro-architecture metrics are under "result" directory. We provide processing scripts for computing the result and plot the figures. Please refer to "README" file for the details.

\subsubsection{Process the metric data and plot the figures}

For E-commerce Intelligence, the processing script is log\_process.sh in the ``aibench\_scenario/ecommerce\_search" directory. The usage of log\_process.sh is ./log\_process.sh \textless log\_dir \textgreater, for example, ./log\_process.sh ../log-0310-1402, the result.csv will be outputted in current directory, and run the command ``python e\_commerce\_search\_figure.py" will produce the Fig.5(1).

For Translation Intelligence, in the ``aibench\_scenario/translation" directory, run the ``python translation\_intelligence\_fig.py" will produce the Fig.5(2).

For the experiment of Fig.7, users have to change the data dimension, recompile the benchmark, and re-run the E-commerce Intelligence benchmark for several times. 

i. Change the data dimension 

Change the Line \#204 of the file 

``aibench\_scenario/framework/online/non\_AI\_module/recommender/cpu\_app.py":

\emph{img = image.img\_to\_array(image.load\_img(img\_path, target\_size=(8, 8)))}

Change (8, 8) to the other dimension you need to test, like (16, 16), (32, 32), (64, 64), (128, 128), or (256, 256).

ii. Recompile the benchmark 

\#cd aibench\_scenario/benchmarks/ecommerce\_search 

\#docker-compose up -d

After changing the data dimension and recompiling the benchmark, re-run the E-commerce Intelligence benchmark according to the steps of "Run E-commerce Intelligence" and obtain the results.

iii. Use the image\_dimension\_figure.sh to produce the Fig.7.

\#./image\_dimension\_figure.sh \textless 8\_logdir\textgreater \textless 16\_logdir\textgreater \textless 32\_logdir\textgreater \textless 64\_logdir \textgreater \textless 128\_logdir\textgreater \textless 256\_logdir\textgreater

Note that the \textless x\_logdir\textgreater are the log directories generated in Step iv when the data dimension is x*x.

For the experiment of Fig.8, lsdata-pact21-fig8.py can produce the corresponding figure, the usage is python  lsdata-pact21-fig8.py \textless input\_dir\textgreater \textless output\_name\textgreater, the input\_dir refer the hw\_log, for example python lsdata-pact21-fig8.py log-0310-1402/hw\_log output. This command will produce an excel file named \textless output\_name\textgreater.xls and a figure. Note that the use\_perf\_new1.py we provide only suit for sampling micro-architecture metrics on Intel Xeon E5645 processors. If you need to profile on another processor type, you need to modify the corresponding hardware performance counters.

% We provide processing scripts and figure plotting scripts to generate the figures used in the paper. Note that the sampling results are saved under "result" directory when test finished.

% i. Compute the performance data and save them in an excel file.

% \#python lsdata.py result result\_new 1

% Parameter "result" means the input directory which contains the sampling results; Parameter "result\_new" means the output excel file name and the output file is result\_new.xls.

% ii. Plot the figures and save them as png image format

% \#python plot.py result\_new.xls

% Parameter "result\_new.xls" is the excel file generated by the first step. After running the command, several png files will be generated. In addition, "pact-AE.txt" is generated for linkage distance analysis.

% iii. Linkage distance computing

% \#\$pact2018/Linkage-Distance

% \#python hiclust\_wiht\_newpca.py pact-AE.txt

% Parameter "pact-AE.txt" is the text file generated by the second step. After running the command, a png file will be generated under the Linkage-Distance directory, which is used as Figure 12 in our paper.

%%%%%%%%%%%%%%%%%%%%%%%%%%%%%%%%%%%%%%%%%%%%%%%%%%%%%%%%%%%%%%%%%%%%%
\subsection{Evaluation and expected result}

{\em To evaluate the scenario benchmarks, users need to run and profile them. The results should follow the similar trends as reported in the paper.
Note that if you download the artifacts and deploy them in your cluster, the results may different from the values in the paper, due to the differences of cluster scale, node configurations, network configurations, etc.}

%%%%%%%%%%%%%%%%%%%%%%%%%%%%%%%%%%%%%%%%%%%%%%%%%%%%%%%%%%%%%%%%%%%%%
\subsection{Experiment customization}

Users can run these scenario benchmarks for different benchmarking purpose, e.g. datacenter deployment strategies. The scenario benchmarks can be deployed on different processors and cluster scales. Also, the AI related TensorFlow serving components can be deployed on different AI chips like GPU and TPU.

%%%%%%%%%%%%%%%%%%%%%%%%%%%%%%%%%%%%%%%%%%%%%%%%%%%%%%%%%%%%%%%%%%%%%
\subsection{Notes}

For the artifact evaluation, it may cost a lot of time to run some benchmarks, we also provide the scripts and the original running logs and profiling data used in our paper, which are suit for our experimental environments and configurations. Since the platform configurations, network configurations, and hardware types (e.g. memory capacity, processing capacity) may be different, so the performance data may be different on another platform.

\end{document}